%% file: ChapBeamTracking_RF2023.tex
\documentclass{cernyrep}
\usepackage[T1]{fontenc}
\usepackage{booktabs}
\usepackage{subfigure,wrapfig}
\usepackage{caption}
\usepackage[bookmarks, colorlinks=true, linktoc=page, pdftex, linkcolor=black, citecolor=black, urlcolor=blue]{hyperref}

\usepackage{fancyhdr}
\fancyhfoffset{4 mm}
\fancypagestyle{ARTTITLE}{%
\fancyhf{} 
\lhead{\hfill CERN Accelerator School Proceedings ---
{\it RF for Accelerators} ---  Berlin, Germany,  2023\hfill}
\lfoot{\hspace{3mm} Available online at \url{https://cas.web.cern.ch/previous-schools}}
\rfoot{\thepage\hspace*{3mm}}
 
}

\usepackage{varwidth}
\usepackage{xcolor}
\begin{document}

\title{Beam Tracking I \& II}
\author{H.~Timko}
\institute{CERN, Geneva, Switzerland}

\begin{abstract}
    Longitudinal beam dynamics simulations are an essential tool for the design and operation of circular and linear accelerators alike, fundamentally supporting everything from RF system design to the interpretation of beam-based measurements. This paper gives a practical, self-contained introduction to longitudinal beam tracking, aimed at newcomers to the field who wish to build, or better understand, their simulation tools. Starting from the analogy between beam tracking and particle-in-cell plasma simulations, the basic equations of motion, their discretisation, and the choice of phase-space coordinates are introduced, followed by the collective effects driven by the machine impedance. The modelling of the RF system and its associated control loops, and the generation and matching of realistic particle distributions are discussed thereafter. Selected topics that require a fully six-dimensional treatment are briefly touched upon, before closing with practical guidance on the design, optimisation, and benchmarking of longitudinal tracking codes.
\end{abstract}

\keywords{Longitudinal beam dynamics, beam tracking, RF systems and feedbacks, \\ collective effects, high-performance computing.}
\maketitle
\thispagestyle{ARTTITLE}

\section{Introduction}
\label{sec:introduction}

    Numerical simulation of the longitudinal beam dynamics is, today, a standard part of the design and operation of any circular or linear accelerator. Such simulations are used to design RF systems and their associated feedback loops, to plan and interpret beam manipulations such as bunch splitting or slip stacking, to predict and mitigate beam instabilities driven by machine impedance, and to support the~interpretation of beam-based measurements more generally. A wide range of dedicated simulation suites has been developed for this purpose across the accelerator community, differing in scope, numerical methods, and level of detail, but sharing a largely common set of underlying concepts.

    This paper collects these common concepts into a practical, self-contained introduction to longitudinal beam tracking, aimed primarily at readers who are new to the field and wish to either use an existing tracking code more effectively, or build their own. The paper follows the structure of the corresponding CAS lectures, and largely follows the conventions and equations used~\cite{Timko2023} in the BLonD synchrotron simulation suite~\cite{BLonD}, developed and maintained at CERN, without being tied to any particular software implementation. Section~\ref{sec:tracking-basics} introduces the basic concepts of beam tracking starting from particle-in-cell simulations, before Section~\ref{sec:longitudinal-tracking} derives the longitudinal equations of motion used throughout the paper. Section~\ref{sec:intensity-effects} discusses collective effects driven by the machine impedance, and Section~\ref{sec:rf-effects} the modelling of the RF system itself, including its associated global and local control loops. Section~\ref{sec:distributions} covers the generation and matching of realistic particle distributions, while Section~\ref{sec:6d-effects} discusses selected effects that require a fully six-dimensional treatment. Finally, Section~\ref{sec:code-design} closes with practical guidance on the design, optimisation, and benchmarking of a tracking code, before some concluding remarks are given in Section~\ref{sec:conclusion}.

\input{text/tracking_basics}

\input{text/longitudinal_tracking}

\input{text/intensity_effects}

\newpage
\input{text/rf_effects}

\newpage
\input{text/particle_distributions}

\newpage
\input{text/6d_effects}

\newpage
\input{text/code_design}

\section{Conclusion}
\label{sec:conclusion}

    These proceedings have given a practical overview of the ingredients for constructing or understanding a~longitudinal beam tracking code: the basic equations of motion and their discretisation; the modelling of collective effects through the machine impedance; the RF system itself, from the cavity-transmitter-beam interaction to the global and local control loops that regulate it; the generation and matching of realistic particle distributions; selected effects that require a fully six-dimensional treatment; and, finally, practical guidance on the design, optimisation, and benchmarking of the resulting code. Naturally, a single paper -- like the two lectures it is based on -- cannot cover every aspect of longitudinal beam dynamics simulation in full depth; the references throughout this paper point to more specialised literature for readers wishing to go further on any specific topic. Taken together, however, the material presented here should provide the reader with the basic tools needed to build, or make effective use of, a simulator of longitudinal beam dynamics.

\appendix
\input{text/appendix_slippage}

\input{text/appendix_symplectic}

\end{document}

%% file: text/tracking_basics.tex
\section{Beam tracking basics} 
\label{sec:tracking-basics}

\subsection{From 6D plasma to beam tracking}
    
    In the 1950's, with the appearance of the first powerful enough computers, the simulation of plasma particles in electromagnetic fields started using the particle-in-cell (PIC) method~\cite{Buneman1959,Dawson1962}. In this method, particles are represented by macro-particles that have the same charge-to-mass ratio as real particles. Fields are calculated on a discrete grid in a stationary (Eulerian) frame, while particles are moved in continuous phase space in a co-moving (Lagrangian) frame; many implicit and explicit integrators, as well as field solvers such as finite-difference and finite-element methods, have been developed for this purpose. Collective effects, e.g.\ collisions or space charge, are often treated in PIC codes on a particle-by-particle basis with Monte-Carlo methods. They are essentially $N_m^2$ problems, where $N_m$ is the number of macro-particles.

    A beam is, fundamentally, a single-component plasma dominated by a strong external RF field. The same underlying concept as in PIC codes applies to beam tracking: fields and their effect on the~particles are calculated on a grid, while particles are tracked in continuous phase space. However, simplifications can be introduced for collective effects: as the external RF field is usually the dominant contribution to the particle energy, other effects are often described through impedance sources instead of particle-by-particle interactions, reducing the problem to a convolution of order $N_m\log(N_m)$, as we will see in Section~\ref{sec:intensity-effects}.

\subsection{Accelerator model and tracking structure}

    A beam-tracking code aims to reproduce, as closely as needed for the application at hand, the components of the real accelerator. The transverse plane is commonly parametrised through a lattice description of dipole, quadrupole, and higher-order magnets, e.g.\ using MAD-X~\cite{MADX} or Xsuite~\cite{Xsuite}. The lattice determines the betatron tunes, chromaticities, and Twiss parameters. For longitudinal applications, it also determines the phase slippage, or drift, experienced by particles as they circulate. Transverse tracking codes can, in addition, model collective effects in the transverse plane, such as transverse impedance and its effect on the beam like coupled-bunch phenomena, head-tail motion, etc.

    The longitudinal tracking, which this paper focuses on, determines the evolution of the RF voltage amplitude, phase and frequency, the change of the synchronous energy following a given magnetic ramp, collective effects due to longitudinal impedance, and other effects such as synchrotron radiation -- in short, the evolution of the particle energy through the kick equation (Section~\ref{sec:eom}). Some phenomena, such as intra-beam scattering or electron-cloud effects, cannot be cleanly separated into a purely longitudinal or transverse problem and require a full six-dimensional (6D) treatment instead (see Section~\ref{sec:6d-effects}).

\begin{figure}
      \centering
      \begin{minipage}[b]{.45\linewidth}
        \centering
        \includegraphics[width=0.8\textwidth]{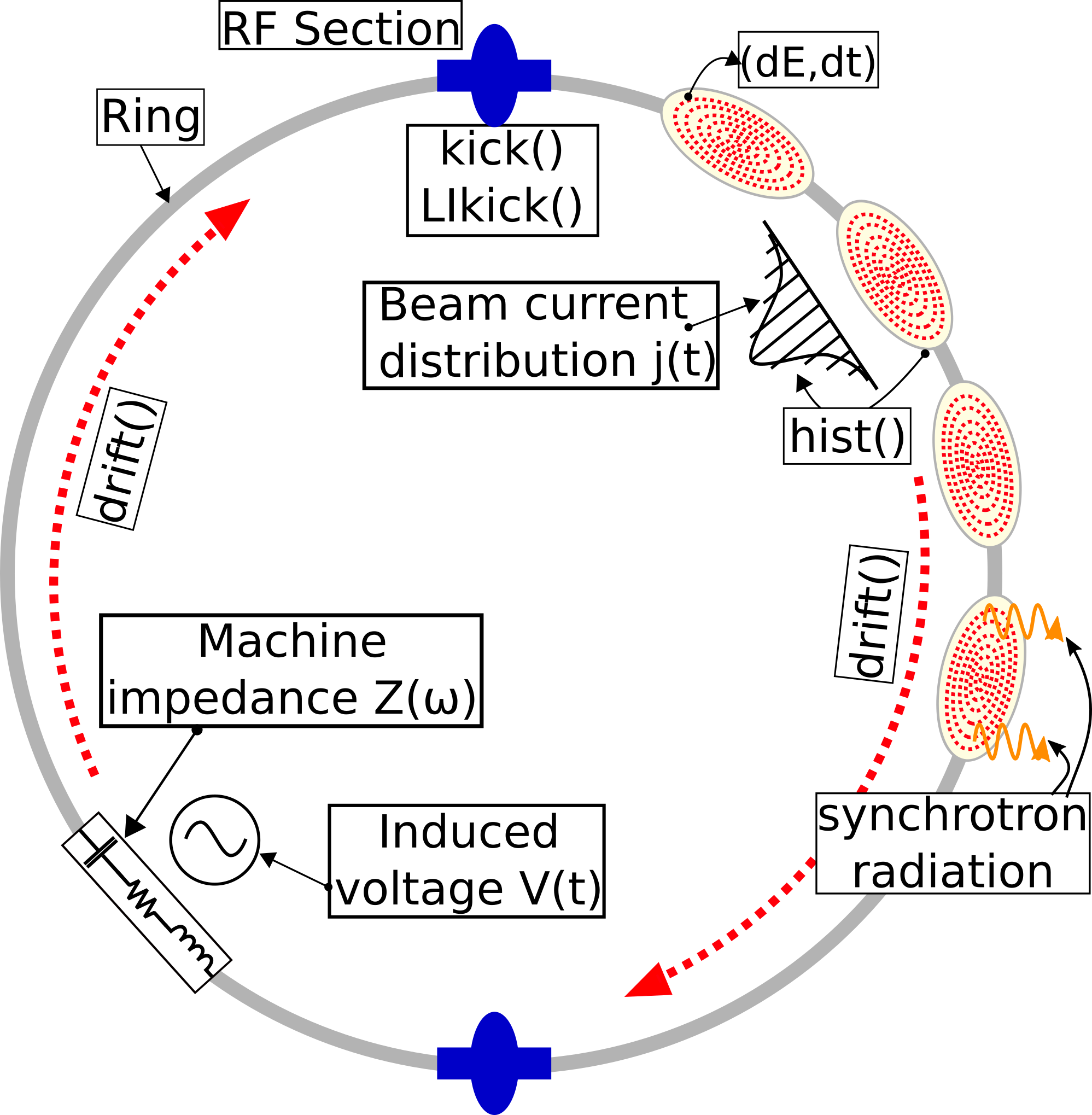}\hfill
        \caption{Example of a synchrotron model, inspired by the BLonD tracking code. Courtesy of K.~Iliakis.}
        \label{fig:blond-ring}
        \vspace{-13pt}
      \end{minipage}
      \hspace{0.05\linewidth}
      \begin{minipage}[b]{.45\linewidth}
        \centering
        \includegraphics[width=0.9\linewidth]{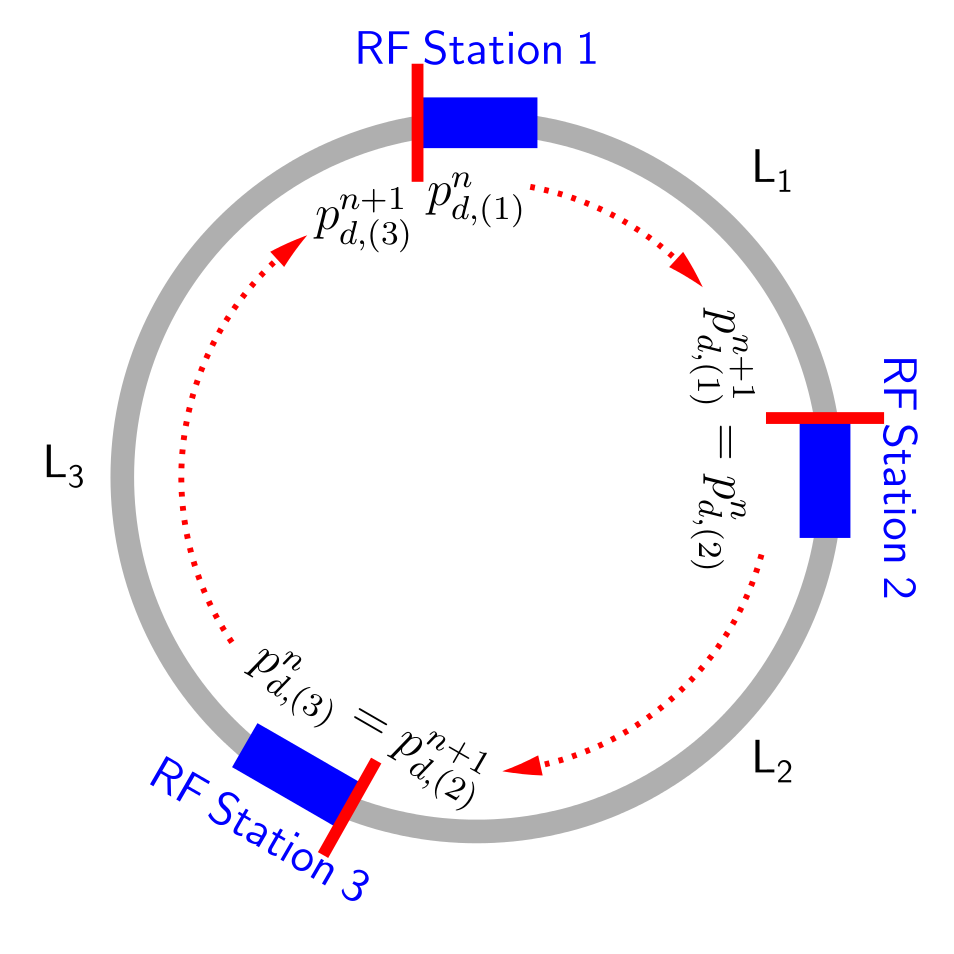}
        \caption{Schematic view of discretisation points around the ring.}
        \label{fig:distretisation}
      \end{minipage}\hfill
    \end{figure}

    In practice, a longitudinal tracking model is built up from a number of interacting components as depicted in Fig.~\ref{fig:blond-ring}: bunch generation (Section~\ref{sec:distributions}), the basic kick-and-drift equations of motion for one or several RF stations providing the RF voltage kick (Section~\ref{sec:eom}), the interaction with the machine impedance through the beam line-density and induced-voltage calculation (Section~\ref{sec:intensity-effects}), and other effects on the particle energy such as synchrotron radiation. Depending on the machine, several such RF stations, each with their own local set of parameters, may need to be modelled along the ring. In this case, discretisation (Section~\ref{sec:discretisation}) is performed on a section-by-section basis, as shown in Fig.~\ref{fig:distretisation}, rather than turn-by-turn.

    A brief remark is in order for linear accelerators, which are also RF machines and whose RF potential and induced voltage can be described in an analogous manner. In a linac, however, a given particle passes through each RF structure only once, so no turn-by-turn tracking is performed. Instead, the model consists of a mapping from one location in the machine to the next, governed by equations of motion different from the synchrotron case. While in a synchrotron the RF cavities can be treated as point-like objects and the particle's energy and velocity change little over a single turn, in a linac, the~finite time needed to traverse a cavity, and the correspondingly large energy gain per structure, must be taken into account. The particle's arrival time with respect to the RF wave is accounted for in both cases.

\subsection{Coordinate choice and equations of motion}
\label{sec:eom}

    In the longitudinal plane, various conjugate coordinate pairs can be used, depending on what is most practical for a given application; an overview can be found in Ref.~\cite{Lee2019}. A classical choice is $(\varphi, \Delta E/\omega_\mathrm{rev})$, where $\varphi$ is the phase of the particle w.r.t.\ the RF wave, $\Delta E = E - E_d$ is the particle's energy relative to the design (synchronous) energy $E_d$, and $\omega_\mathrm{rev}$ is the angular revolution frequency. Alternatively, the~pair $(\varphi, \delta)$ can be used, with the~relative momentum offset $\delta \equiv \Delta p/p_d \simeq \Delta E/(\beta_d^2 E_d)$, where $\beta_d$ is the~relative velocity corresponding to $E_d$. With these coordinates, the phase or drift equation becomes~\cite{Lee2019}
    \begin{equation}
        \dot \varphi = \frac{h \omega_\mathrm{rev}^2 \eta}{\beta_d^2 E_d} \left( \frac{\Delta E}{\omega_\mathrm{rev}} \right) = h \omega_\mathrm{rev} \eta \delta \, ,
        \label{eq:drift-coord1}
    \end{equation}
    where $h$ is the harmonic number of the RF system and $\eta = \eta(\delta) = \eta_0 + \eta_1 \delta + \eta_2 \delta^2 + ...$ is the slippage factor, see also Appendix~\ref{app:slippage}. The energy, or kick, equation becomes
    \begin{equation}
        \frac{d}{d t} \left( \frac{\Delta E}{\omega_\mathrm{rev}} \right) = \frac{\beta_d^2 E_d}{\omega_\mathrm{rev}} \dot \delta = \frac{e V}{2 \pi} (\sin{\varphi} - \sin{\varphi_d}) \, ,
        \label{eq:kick-coord1}
    \end{equation}
    where $V$ is the RF voltage amplitude and $\varphi_d$ is the phase of the synchronous particle. The Hamiltonian for this system of equations of motion is then
    \begin{equation}
        H \left( \varphi, \frac{\Delta E}{\omega_\mathrm{rev}} \right) = \frac{1}{2} \frac{h \eta \omega_\mathrm{rev}}{\beta_d^2 E_d} \left( \frac{\Delta E}{\omega_\mathrm{rev}} \right)^2 + \frac{e V}{2 \pi} \{\cos{\varphi} - \cos{\varphi_d} + (\varphi - \varphi_d) \sin{\varphi_d} \}
        \label{eq:hamiltonian-coord1}
    \end{equation}
    or, equivalently,
    \begin{equation}
        H \left( \varphi, \delta \right) = \frac{1}{2} h \eta \omega_\mathrm{rev} \delta^2 + \frac{\omega_\mathrm{rev} e V}{2 \pi \beta_d^2 E_d} \{\cos{\varphi} - \cos{\varphi_d} + (\varphi - \varphi_d) \sin{\varphi_d} \} \, .
        \label{eq:hamiltonian-coord2}
    \end{equation}

    In the remainder of this paper, we will use the pair $(\Delta t, \Delta E)$, where $\Delta t = t - t_d$ is the arrival time of the particle at the RF cavity w.r.t.\ the arrival time $t_d$ of the design (synchronous) particle; the~corresponding equations of motion are given in Section~\ref{sec:longitudinal-tracking}.

    For code-optimisation purposes, the use of dimensionless coordinates can be considered. To re-scale physical variables to dimensionless ones, the typical scale of a given unit in the system can be used, e.g.\ $E \rightarrow E/E_d$. Sometimes numerical factors have to be taken into account explicitly: for instance, the~kick equation requires the evaluation of trigonometric functions, which is usually optimised for a phase range of $(-\pi,\pi)$ or $(0, 2\pi)$. Therefore, whenever the time coordinate is re-scaled, e.g.\ as $t \rightarrow t/t_d$, wrapping with $2\pi$ should be taken into account for the corresponding phase.

\subsection{Discretisation}
\label{sec:discretisation}

    Beam tracking codes advance the particle coordinates turn by turn. Within a single turn, the energy kick imparted by an RF cavity, Eq.~\eqref{eq:kick-coord1}, is applied assuming the cavity to be point-like, i.e.\ the kick is integrated over the passage through the cavity and is otherwise treated as instantaneous; the time or phase drift over the remainder of the ring, Eq.~\eqref{eq:drift-coord1}, instead relies on an average slippage factor, as obtained from lattice calculations.

    \begin{wrapfigure}{r}{0.45\textwidth}
      \centering
      \vspace{-18pt}
      \includegraphics[width=0.43\textwidth]{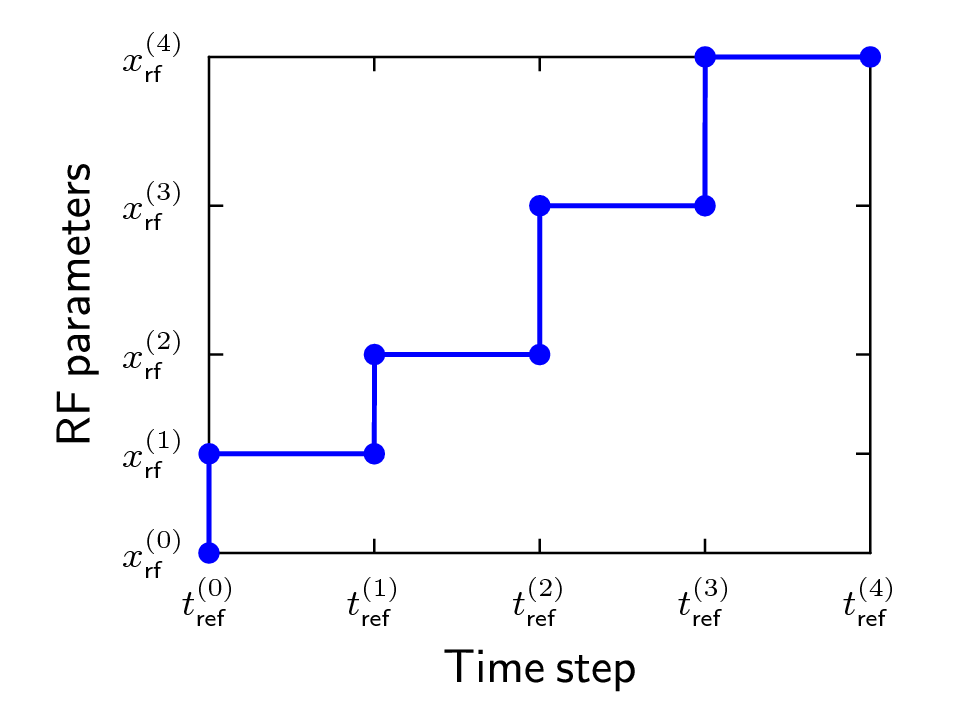}
      \caption{Example: step-wise discretisation of RF parameters.}
      \vspace{-12pt}
      \label{fig:discretisation-steps}
    \end{wrapfigure}    
    In some synchrotrons, the change of the particle coordinates within a single turn is not negligible; for instance, in machines with strong synchrotron radiation damping, such as future circular colliders (FCC). In such cases, the modelling of several RF stations and a sub-cycling of the kick-and-drift sequence may be needed, applying several such sequences per turn. In this case, the impedance and the slippage factor can also differ from one section of the ring to another, and need to be tracked accordingly.

    Beyond the turn-by-turn discretisation of the particle coordinates themselves, any time-dependent machine or beam parameter (RF voltage, phase, frequency, impedance tables, etc.) needs to be discretised as well. Depending on the application and the expected rate of change of the given quantity, this can be done through a step-wise constant approximation (see Fig.~\ref{fig:discretisation-steps}), updated once per turn or once per given number of turns, or through interpolation between sampled input points.

%% file: text/longitudinal_tracking.tex
\section{Longitudinal tracking}
\label{sec:longitudinal-tracking}

    In the following, the concrete equations are based on the Beam Longitudinal Dynamics (BLonD) simulation suite~\cite{BLonD}.

\subsection{Reference frame}

    A tracking code has to relate the beam-particle coordinates, and the RF cavity voltage, phase and frequency, to some common time base -- exactly as beam- and cavity-control systems do in a real machine, by referring everything to an external clock. This external reference determines a design clock time $t_{d,(n)}$, obtained at turn $n$ by accumulating the design revolution periods of every preceding turn,
    \begin{equation}
      \label{eq:referenceTime}
      t_{d,(0)} \equiv 0 \phantom{xx}\text{and}\phantom{xx} t_{d, (n)} \equiv \sum_{k=1}^n T_{\mathrm{rev},(k)} \phantom{x}\mathrm{for}\phantom{x} n \geq 1 .
    \end{equation}
    Each revolution period follows from the radius of the design orbit $R_d$ and from $\beta_{d,(n)}$, the design particle's velocity on that orbit relative to the speed of light $c$,
    \begin{equation}
      \label{eq:revolutionPeriod}
      T_{\mathrm{rev},(n)} = \frac{2 \pi R_d}{\beta_{d,(n)}c} \, .
    \end{equation}
    For the turns to be simulated, either $\beta_{d,(n)}$ has to be input directly, or the user may choose to prescribe either the design momentum $p_{d,(n)}$ or the design total energy $E_{d,(n)}$. In all cases, this input determines the required magnetic field $B_{d,(n)}$ at each turn through
    \begin{equation}
      \label{eq:magneticField}
      p_{d,(n)} = |q| \rho B_{d,(n)} \, ,
    \end{equation}
    with $q$ being the particle charge and $\rho$ the bending radius of the magnets.

\subsection{Equations of motion}

    Each particle is tracked using the phase-space coordinates $(\Delta t_{(n)}, \Delta E_{(n)})$: its arrival time relative to the~accumulated reference time $t_{d,(n)}$, and its energy relative to the design total energy $E_{d,(n)}$. Within a~given turn or ring section, the particle's energy is updated first, from iteration $n$ to iteration $n+1$, by summing the contributions of every RF voltage kick delivered by that section's RF station, each evaluated at the~particle's own arrival time,
    \begin{align}
      \Delta E_{(n+1)} &=  \Delta E_{(n)} + \sum_{k=1}^{n_{\mathrm{rf}}} q V_{k,(n)} \sin(\omega_{\mathrm{rf},k,(n)} \Delta t_{(n)} + \varphi_{\mathrm{rf},k,(n)}) - \nonumber \\
      &- (E_{d,(n+1)} - E_{d,(n)}) + E_{\mathrm{other},(n)} \, ,
      \label{eq:kick}
    \end{align}
    with $V_k$, $\omega_{\mathrm{rf},k}$, and $\varphi_{\mathrm{rf},k}$ denoting, respectively, the voltage amplitude, angular frequency, and phase of RF system $k$, and $E_{d,(n+1)} - E_{d,(n)}$ subtracting out the turn-by-turn change of the design energy itself. Any remaining energy change -- from induced voltage, synchrotron radiation, or similar sources -- is collected in the last term, $E_{\mathrm{other},(n)}$.

    Once the energy has been updated, the arrival time follows next, using the momentum compaction factor $\alpha$, taken here to at least zeroth and up to second order,
    \begin{align}
      \Delta t_{(n+1)} &= \Delta t_{(n)} + T_{\mathrm{rev},(n+1)}\Bigg[  \Big( 1 + \alpha_{0,(n+1)} \delta_{(n+1)}  + \nonumber \\
      &+ \alpha_{1,(n+1)} \delta^2_{(n+1)} + \alpha_{2,(n+1)} \delta^3_{(n+1)} \Big) \frac{1 + \frac{\Delta E_{(n+1)}}{E_{d,(n+1)}}}{1 + \delta_{(n+1)}} - 1 \Bigg], \ ,
      \label{eq:drift}
    \end{align}
    with the relative momentum offset given by $\delta_{(n)} = \frac{\Delta p_{(n)}}{p_{d,(n)}}\simeq\frac{\Delta E_{(n)}}{\beta_d^2 E_{d,(n)}}$. The relationship of the slippage and momentum compaction factor is detailed in Appendix~\ref{app:slippage}.

\subsection{Symplecticity}

    The term $E_{\mathrm{other},(n)}$ in Eq.~(\ref{eq:kick}) may generally depend on the particle's own coordinates, $E_{\mathrm{other},(n)}(\Delta t_{(n)}, \Delta E_{(n)})$. Setting it aside for a moment, Eqs.~(\ref{eq:drift}) and (\ref{eq:kick}) reduce to a one-turn map that transforms $\mathbf{x} \equiv (\Delta t_{(n)}, \Delta E_{(n)})$ to $\mathbf{y} \equiv (\Delta t_{(n+1)}, \Delta E_{(n+1)})$, via
    \begin{align}
        \Delta t_{(n+1)} &= \Delta t_{(n)} + f\left(\Delta E_{(n+1)}(\Delta t_{(n)},\Delta E_{(n)})\right)\,,
        \label{eq:drift_abstract}\\
        \Delta E_{(n+1)} &= \Delta E_{(n)} + g\left(\Delta t_{(n)}\right) \label{eq:kick_abstract}\,.
    \end{align}
    Applying the chain rule, the corresponding Jacobian, $\mathcal{M}_{ij} \equiv \partial y_{i}/\partial x_{j}$, takes the form
    \begin{equation}
        \mathcal{M} =
        \begin{pmatrix}
        1 + f' g' & f' \\
        g' & 1
        \end{pmatrix} \, .
        \label{eq:mapping-matrix}
    \end{equation}
    This matrix can readily be shown to obey the symplectic condition~\cite{wolski}
    \begin{equation}
        \label{eq:simplectic_condition}
        \mathcal{M}^\mathrm{T} \mathcal{S} \mathcal{M} = \mathcal{S}\,,
    \end{equation}
    with $\mathcal{S} \equiv \begin{pmatrix}
        0 & 1\\
        -1 & 0
    \end{pmatrix}$.
    Consequently, in the absence of collective effects, the bare equations of motion conserve the phase-space area enclosed by the particle trajectories, even with the fairly involved form of Eq.~(\ref{eq:drift}). This symplectic property can be, however, lost once collective effects are folded into the tracking.

    The general derivation of the symplectic condition used above, for an arbitrary Hamiltonian system of conjugate coordinates, is given in Appendix~\ref{app:symplectic}.

\subsection{Synchronism}

    The condition for a particle to remain \emph{synchronous} when tracked from time step $n$ to $n+1$ is to enter and exit the cavity with exactly the design energy, $\Delta E_{(n)} = \Delta E_{(n+1)} = 0$, or equivalently $E_{(n)} = E_{d,(n)}$ and $E_{(n+1)} = E_{d,(n+1)}$. As a consequence, the kick equation for this particle reduces to
\begin{equation}
    \sum_{k=1}^{n_{\mathrm{rf}}} q V_{k,(n)} \sin(\omega_{\mathrm{rf},k,(n)} \Delta t_{(n)} + \varphi_{\mathrm{rf},k,(n)}) - (E_{d,(n+1)} - E_{d,(n)}) + E_{\mathrm{other},(n)} = 0
\end{equation}
and the drift equation reduces to
\begin{equation}
    \Delta t_{(n+1)} = \Delta t_{(n)} + 0 \, .
\end{equation}

Consider now the simple example of longitudinal motion without intensity effects and with constant RF parameters. In two consecutive turns, the kick equations for the synchronous particle read as
\begin{equation}
    \sum_{k=1}^{n_{\mathrm{rf}}} q V_{k} \sin(\omega_{\mathrm{rf},k} \Delta t_{(n)} + \varphi_{\mathrm{rf},k}) = E_{d,(n+1)} - E_{d,(n)}
\end{equation}
and 
\begin{equation}
    \sum_{k=1}^{n_{\mathrm{rf}}} q V_{k} \sin(\omega_{\mathrm{rf},k} \Delta t_{(n+1)} + \varphi_{\mathrm{rf},k}) = E_{d,(n+2)} - E_{d,(n+1)} \, .
\end{equation}
If the acceleration rate is not constant, i.e.\ $E_{d,(n+1)} - E_{d,(n)} \neq E_{d,(n+2)} - E_{d,(n+1)}$, then $\Delta t_{(n)} \ne \Delta t_{(n+1)}$, meaning that the particle that was synchronous from $n \rightarrow n+1$ is no longer synchronous from $n+1 \rightarrow n+2$.

\subsection{Periodicity}

    By construction of the turn-by-turn discretisation introduced above, the time coordinate of any particle at a given turn $n$ should lie within $(0, T_{\mathrm{rev},(n)})$. In some cases, however, particles can cross these time boundaries -- for instance, in machines with a small harmonic number $h$, when tracking a full machine, or in the presence of significantly debunched beam. To handle this correctly, periodic boundary conditions have to be introduced explicitly into the coordinate frame: particles with $\Delta t_{(n)} < 0$ are lagging behind and have to be tracked twice within a single call to be in the same time frame as the other particles; conversely, particles with $\Delta t_{(n)} > T_{\mathrm{rev},(n)}$ have already reached the next turn and have to be put on hold for a turn before being tracked further. 
    
    A practical example of such periodicity handling is the simulation of the beam-injection process into the CERN PS Booster (PSB), where incoming particles must be correctly folded into the existing turn structure of the ring, whether the RF frequency program is locked to the magnetic-field ramp or kept at a fixed frequency.

\subsection{RF gymnastics}

    In machines with multi-harmonic RF systems, a sequence of RF manipulations -- colloquially referred to as \emph{RF gymnastics} -- can be applied by programming the RF voltage of the individual harmonics as a function of time. Common examples include (i) splitting or merging of bunches, achieved by adiabatically ramping up (or down) a higher-harmonic RF voltage relative to the fundamental; (ii) bunch rotation, achieved by non-adiabatically increasing the RF voltage and then extracting or recapturing the bunch after a quarter of a synchrotron period; and (iii) bunch compression, achieved by changing the harmonic number of a given RF system while keeping the beam bunched throughout. 
    
    A well-known example combining several such manipulations is the bunch-compression, merging, and splitting (BCMS) scheme used in the CERN PS, which is also used to prepare bunches for double splitting and bunch rotation ahead of capture in the SPS.

    Longitudinal painting, to name another example, is used to deliberately fill a chosen part of the~longitudinal phase space, typically because the injected emittance would otherwise be too small for the~longitudinal stability requirements of the machine the beam is injected into. A common implementation is multi-turn injection, in which small-emittance bunches are injected at slightly different energy offsets over a number of successive turns, such that the final, composite bunch fills the full bucket -- as is done, for example, in the PSB. From a computational point of view, this requires the data structure holding the beam coordinates to be dynamically expanded turn by turn as new particles are injected, rather than being fixed at the start of the simulation.

%% file: text/intensity_effects.tex
\section{Intensity and energy effects}
\label{sec:intensity-effects}

    In most beam-tracking applications, particle-particle interactions are handled with a particle-in-cell-like solver~\cite{BirdsallLangdon} rather than a direct sum over all pairs. To do so, the beam is binned onto a grid, and the resulting induced voltage $V_\mathrm{ind}(\Delta t)$ is evaluated cell by cell as the convolution of the beam line density $\lambda(t)$ with the wake function $W(t)$,
    \begin{equation}
        V_\mathrm{ind}(\Delta t) = - q\, N_p\int_{-\infty}^{+\infty} \lambda(\tau) \, W(\Delta t-\tau) \, d \tau  \, ,
        \label{eq:Vind_time}
    \end{equation}
    with the line density normalised so that $\int_{-\infty}^{+\infty} \lambda(t)\,dt = 1$ and $N_p$ being the number of real particles represented by the beam. The overall sign is chosen such that a positive wake potential corresponds to an energy loss. This induced voltage feeds back into the tracking equations as an additional energy kick, $E_\mathrm{ind}(\Delta t) = q\, V_\mathrm{ind}(\Delta t)$, summed into the $E_{\mathrm{other}}$ term of the kick equation, Eq.~\eqref{eq:kick}, and is evaluated ahead of the drift step, should the energy be updated first.

    Equivalently, the induced voltage can be obtained in the frequency domain, via the impedance $Z(\omega)$, itself the Fourier transform of the wake function,
    \begin{equation}
        Z(\omega) = \int_{-\infty}^{+\infty} W(t) \, e^{-j \, \omega \, t} \, d t \, ,
    \end{equation}
    so that the induced voltage is recovered as the inverse transform of the impedance multiplied by the beam spectrum $\Lambda(\omega)$, the latter being the Fourier transform of the line density itself,
    \begin{equation}
        V_\mathrm{ind}(\Delta t) = - \frac{ q\, N_p}{2\,\pi} \int_{-\infty}^{+\infty} Z(\omega) \,  \Lambda(\omega)\,e^{j \, \omega \, \Delta t} \, d\Delta t \, .\label{eq:Vind_freq}
    \end{equation}
    The two formulations are mathematically equivalent; which one is more practical depends on the discretisation required for the numerical simulation at hand, and hence on the bandwidth of the impedance source under consideration.

\subsection{Discretising the induced voltage}

    Once the beam profile is discretised on an equidistant grid, $\lambda[m] \equiv \lambda(t_m)$, the induced voltage can be evaluated in the frequency domain, given the machine impedance $Z[i]$, as
    \begin{equation}
        V_\mathrm{ind}[k]  = - q \, N_p \, \mathrm{IFFT}\left( Z[i] \ \Lambda [i]\right) = - q \, N_p \, \mathrm{IFFT}\left( Z[i] \ \mathrm{FFT}(\lambda [m])\right) \, ,
        \label{eq:imp-freq-discr}
    \end{equation}
    where the multiplication is computationally cheaper than an explicit convolution, at the cost of having to compute the beam spectrum via a forward and inverse Fourier transform. Alternatively, in the time domain, a discrete convolution can be evaluated directly,
    \begin{equation}
        V_\mathrm{ind}[k] = - q\, N_p\ \sum_m\nolimits\lambda[m] \, W[k-m]  \, ,
        \label{eq:imp-time-discr-conv}
    \end{equation}
    though unless small arrays are used, this has a computational complexity of $\mathcal{O}(K\times M)$ and is correspondingly slow. A faster time-domain approach again relies on FFTs, via the circular convolution theorem,
    \begin{equation}
        V_\mathrm{ind}[k]  = - q \, N_p \, \mathrm{IFFT}\left\{ \mathrm{FFT}\left( W[n] \right) \, \mathrm{FFT}\left( \lambda [n]\right)\right\} \, ,
        \label{eq:imp-time-discr}
    \end{equation}
    which reduces the complexity to $\mathcal{O}(N\log N)$, provided that careful zero-padding is applied to ensure $N \geq K+M-1$, so that the result corresponds to a genuine linear (rather than circular) convolution. In all cases, it is advisable to rely on a well-optimised FFT library, such as FFTW~\cite{FFTW}, rather than re-implementing the transform.

\subsection{Impedance sources}

    Impedance models can be imported as tables, e.g.\ based on the electromagnetic modelling of accelerator elements with a tool such as CST Studio~\cite{CST}, or described by dedicated analytical models for specific machine elements. A resonator, commonly used to represent narrow-band structures such as superconducting cavities, is described in the frequency domain by
    \begin{equation}
        Z(\omega) = \frac{R_s}{1 + j Q \left(\frac{\omega}{\omega_r}-\frac{\omega_r}{\omega}\right)}
        \label{eq:imp-res-freq}
    \end{equation}
    and in the time domain by
    \begin{align}
        W(t>0) &= 2\alpha R_s e^{-\alpha t}\left(\cos{\tilde{\omega}t} - \frac{\alpha}{\tilde{\omega}}\sin{\tilde{\omega}t}\right) \\
        W(0) &= \alpha R_s \, , \textrm{ with } \alpha = \frac{\omega_r}{2Q} \textrm{ and } \tilde{\omega} = \sqrt{\omega_r^2 - \alpha^2} \, ,
        \label{eq:imp-res-time}
    \end{align}
    where $R_s$ is the shunt impedance, $Q$ the cavity quality factor, and $\omega_r$ the resonant frequency. Note that using a very low $Q$ of the order of $\mathcal{O}(1)$, also broad-band impedances can be modelled.
    
    A travelling-wave cavity~\cite{Dome1977} is instead described in the frequency domain by
    \begin{align}
        Z &= Z_+ + Z_- \textrm{   where   } \\
        Z_\pm(f) &\equiv R_s \left[\left(\frac{\sin{\frac{\tilde{\omega}_\pm\tau}{2}}}{\frac{\tilde{\omega}_\pm\tau}{2}}\right)^2 \mp 2i \frac{\tilde{\omega}_\pm \tau - \sin{ \tilde{\omega}_\pm \tau}}{\left( \tilde{\omega}_\pm \tau \right)^2}\right]  \textrm{   and   } \tilde{\omega}_\pm \equiv \omega \pm \omega_r
        \label{eq:imp-twc-freq}
    \end{align}
    and in the time domain by
    \begin{equation}
        W(t) =
        \begin{cases}
            \frac{4R_s}{\tau}\left(1-\frac{t}{\tau}\right)\cos{\omega_r t},& \textrm{if } 0<t<\tau\\
            \frac{2R_s}{\tau} ,              & \textrm{if } t=0 \, ,
        \end{cases}
        \label{eq:imp-twc-time}
    \end{equation}
    where $\tau$ is the cavity filling time. 
    
    Resistive-wall impedance, relevant for beam-pipe sections of length $L$ and radius $b$, is modelled in the frequency domain as
    \begin{equation}
        Z(f) = \frac{Z_0 c L}{ \pi } \frac{ 1 }{ \left[1 - i\, \mathrm{sgn} f \right] 2  b  c
                                        \sqrt{ \frac{\sigma_c Z_0 c }{ 4 \pi |f| } + i 2 \pi b^2 f } } \, ,
        \label{eq:imp-rwi-freq}
    \end{equation}
    where $Z_0$ is the vacuum impedance and $\sigma_c$ the conductivity of the pipe wall. 
    
    Finally, a constant $\Im(Z/n)$ is often used to represent, e.g.\ loss of Landau damping or space-charge effects,
    \begin{equation}
        \frac{Z}{n} = \frac{Z}{f/f_\mathrm{rev}} = \textrm{const.}
        \label{eq:imp-ZoN-freq}
    \end{equation}
    This simplifies the computation of the induced voltage considerably, replacing the FFTs by the derivative of the line density,
    \begin{equation}
        V_\mathrm{ind}[k] = -\frac{q\, T_\mathrm{rev}}{2 \, \pi\, T_s} \frac{Z}{n}  \frac{d\lambda [k]}{dn} \, ,
        \label{eq:imp-SC-freq}
    \end{equation}
    which is particularly useful for space-charge simulations, where large numbers of macro-particles might otherwise be required; the caveat is that the derivative of the line density can add numerical noise to the~induced voltage.

\subsection{Resolving beam and impedance}

    The induced voltage is usually added to a simulation in order to look for stable or unstable solutions, but there is no generic recipe for how finely the beam and impedance need to be resolved: what matters is to resolve whatever phenomenon is of interest. For example, microwave instabilities require a fine grid over a single bunch, whereas coupled-bunch instabilities require the long-range wake to be resolved over many bunches or turns.
    
    A practical approach is to design the simulation by comparing the impedance spectrum to the~beam spectrum. Given a desired frequency resolution $\Delta f$ and maximum frequency $f_\mathrm{max}$ to be resolved, the~corresponding time-domain binning follows as $\Delta t_\mathrm{bin} = 1/f_\mathrm{max}$ and $t_\mathrm{max} = N \Delta t_\mathrm{bin} = 1/\Delta f$. Machine impedance can often be decomposed into a sum of resonators, Eq.~\eqref{eq:imp-res-freq}. Two limiting regimes can be distinguished, depending on the relation between the resonator bandwidth $\omega_r/(2Q)$ and the inverse bunch length $1/\tau$: in the broad-band (or short-bunch) regime, the bunch spectrum samples the impedance integral, i.e.\ effectively $R_s/Q$, whereas in the narrow-band (or long-bunch) regime, the bunch spectrum samples the impedance at a single frequency, i.e.\ effectively $R_s$. Impedances that drive instabilities most effectively are those with $\omega_r/(2Q)\sim 1/\tau$.
    
    The choice of binning follows directly from this picture: a long wake or a narrow impedance feature requires a large $t_\mathrm{max}$, or equivalently a small $\Delta f_\mathrm{bin}$, and may require several turns to be resolved simultaneously. Conversely, a short wake or a broad impedance feature requires a small $\Delta t_\mathrm{bin}$, or equivalently a large $f_\mathrm{max}$. Care should also be taken to apply a physically motivated cut-off frequency to broad-band impedance models, since some analytical models produce unphysical impedance at high frequency, which can in turn drive unphysical beam instabilities. Finally, a finer time-domain binning generally requires more simulation macro-particles to keep the statistical noise of the beam profile under control. It is good practice to perform convergence studies in both the number of particles and the number of profile bins before drawing physical conclusions. Typical observables used to characterise beam (in)stability are the dipole (centre-of-mass) and quadrupole (bunch-length) oscillations of the bunch.
    
\subsection{Multi-turn wake}

\begin{figure}[ht]
      \centering
      \includegraphics[width=0.43\textwidth]{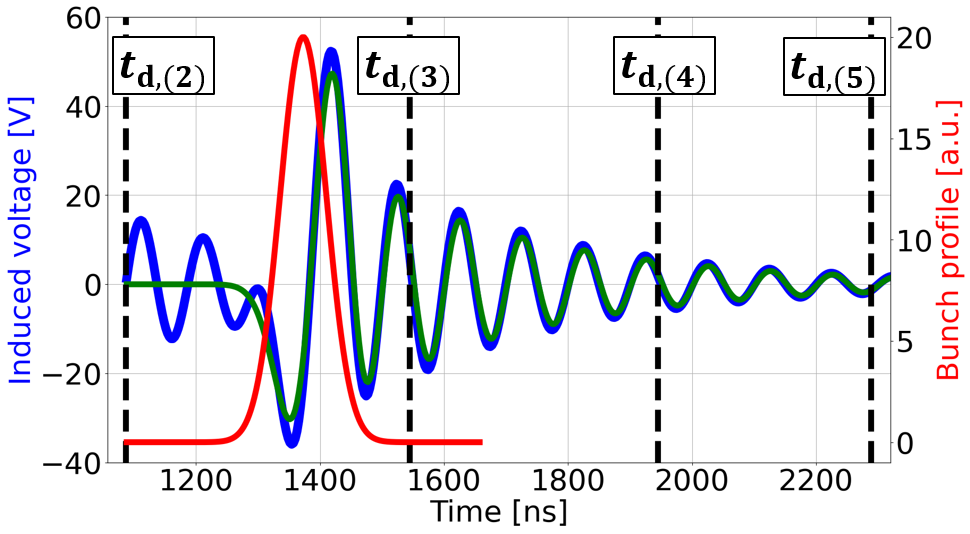}
      \caption{Induced voltage calculated over three machine turns.}
      \vspace{-18pt}
      \label{fig:multi-turn}
      \end{figure}

    Impedances with a long memory can couple particles over several machine turns, as depicted in Fig.~\ref{fig:multi-turn}, in which case the beam profile has to be kept in memory over the corresponding number of past turns in
    order to evaluate Eq.~\eqref{eq:Vind_time} correctly. 
    This can become computationally demanding for machines with long revolution periods and long-range wakes. For instance, multi-turn wake calculations in the CERN PSB may need of the order of 300~turns of beam profile history to be tracked simultaneously.
    
    Several advanced schemes exist to reduce the resulting computational cost. When the bunch spacing is large compared to the bunch length, a compressed convolution can be used, in which only the~relevant, non-zero data points are retained and a window function accounts for the (mostly empty) remainder of the ring~\cite{Komppula2018,Karpov2021}. Alternatively, sparse slicing or a non-uniform grid can be used, relying on a~non-uniform Fourier transform~\cite{Schwarz2021} or on more exotic transforms such as the chirp $z$-transform~\cite{Vadai2021}.

\subsection{Potential-well distortion}

In a first approximation, the induced voltage leads to a shift of the synchronous phase and of the central synchrotron frequency -- an effect known as potential-well distortion. In the single-RF, short-bunch approximation, the perturbation of the potential well can be described as a driven harmonic oscillator,
\begin{equation}
    \frac{d^2 \Delta \varphi}{dt^2} + \omega_{s0}^2 \Delta\varphi = \frac{\omega_{s0}^2}{V_\mathrm{rf}\cos{\varphi_s}} V_\mathrm{ind}(\Delta \varphi) \, .
    \label{eq:pwd-1}
\end{equation}
For a long-range wake field and a beam spectrum that can be taken as constant over the frequency range of interest, the induced voltage on the right-hand side can instead be written as a sum over the impedance sampled at multiples of the revolution frequency~\cite{Laclare1985},
\begin{equation}
    V_\mathrm{ind}(\Delta \varphi) = -\frac{N_p q \omega_\mathrm{rev}}{2\pi}\sum_{p=-\infty}^{\infty} Z(p\omega_\mathrm{rev})\,\Lambda(p\omega_\mathrm{rev})\,e^{jp\omega_\mathrm{rev}t} \, .
    \label{eq:pwd-2}
\end{equation}
Expanding the exponential and the phase around $\varphi_s$ then results in a shift of the synchronous phase of
\begin{equation}
    \Delta \varphi_s = -\frac{N_p q \omega_\mathrm{rev}}{2 \pi V_\mathrm{rf} \cos{\varphi_s}} \sum_{p=-\infty}^\infty \left( \Re(Z(p\omega_\mathrm{rev})) |\Lambda(p\omega_\mathrm{rev}) |^2 \right) \, ,
    \label{eq:pwd-3}
\end{equation}
and a shift in synchrotron frequency of
\begin{equation}
    \frac{\Delta \omega_s}{\omega_{s0}} \approx \frac{N_p q \omega_\mathrm{rev}}{4 \pi V_\mathrm{rf} \cos{\varphi_s}} \Im(Z/n) \sum_{p=-\infty}^\infty \left(  p^2 \Lambda(p\omega_\mathrm{rev}) \right) \, .
    \label{eq:pwd-4}
\end{equation}
In addition, for $\gamma > \gamma_T$ and $\Im(Z/n) > 0$, a bunch lengthening is observed with respect to the zero-intensity bunch length $\tau_0$,
\begin{equation}
    \left(\frac{\tau}{\tau_0} \right)^2 = \frac{\omega_{s0}}{\omega_s} \sqrt{\frac{\cos{\varphi_s}}{\cos{\varphi_s + \Delta \varphi_s}}} \, .
    \label{eq:pwd-5}
\end{equation}

\subsection{Loss of Landau damping}

Landau damping, originally derived for plasmas~\cite{Landau1946}, requires a spread of synchrotron frequencies among the particles of a bunch in order to keep the bunch stable; without such a spread, the bunch would instead oscillate coherently. Keeping for instance the energy, RF voltage, and emittance constant and increasing the bunch intensity, eventually, a coherent oscillation line separates from the continuous, incoherent bunch spectrum. The intensity at which this separation occurs marks the threshold for loss of Landau damping, and is driven by $\Im(Z/n)$. Depending on the impedance source, this can be a~single-bunch or a multi-bunch effect.

In particle tracking, loss of Landau damping can be probed by applying a small phase kick to a~bunch (or beam) that is initially in steady state: if the resulting bunch oscillations fail to damp, Landau damping has been lost. In practice, it can be difficult to define a robust threshold on the oscillation amplitude that distinguishes genuine undamped oscillations from residual background or numerical noise.

\subsection{Synchrotron radiation}

In longitudinal particle tracking, synchrotron radiation can be described through two energy kicks added to Eq.~\eqref{eq:kick},
\begin{equation}
    E_{\mathrm{other},(n)} = - U_0 - \frac{2}{\tau_z} \Delta E_{(n)} \, ,
    \label{eq:SR}
\end{equation}
where the first term on the right-hand side represents the average energy loss of the bunch per turn,
\begin{equation}
    U_0 = \frac{4\pi}{3}\frac{r_\mathrm{cl}}{m_p^3c^6}\frac{1}{\rho}E_{d,(n)}^4\frac{R}{C} \, ,
    \label{eq:SR-ave}
\end{equation}
with $m_p$ being the particle mass, $\rho$ the magnet bending radius, and $r_\mathrm{cl}$ the classical particle radius. The~second term in Eq.~(\ref{eq:SR}) represents the difference in energy loss between individual particles, proportional to their energy offset and characterised by the longitudinal damping time $\tau_z$ per turn. For machines with a~large energy loss per turn, this energy loss should be applied progressively via sub-cycling rather than as a single large discrete kick per turn, in order to avoid unphysical results; sub-cycling of the~synchrotron-radiation kick and drift can be applied even in ring sections without an RF cavity.

For short bunches, quantum excitation must also be taken into account: the statistical nature of photon emission counteracts, in part, the radiation damping, and is described as a random energy kick~\cite{Muller2017},
\begin{equation}
    E_{\mathrm{other},(n)} = 2 \frac{\sigma_{\Delta E}}{\sqrt{\tau_z}} E_{d,(n)} \mathtt{RANDN} \, ,
    \label{eq:QE}
\end{equation}
where $\sigma_{\Delta E}$ is the equilibrium energy spread and $\mathtt{RANDN}$ a normally distributed random number.

Coherent synchrotron radiation, relevant when the radiated wavelength becomes comparable to the bunch length, can also be modelled as an impedance. The free-space model gives~\cite{Murphy1997}
\begin{align}
    &\frac{\Re Z}{Z_0} = \frac{\sqrt{3}\gamma}{4}\frac{f}{f_\mathrm{crit}}
    \int_{f/f_\mathrm{crit}}^\infty K_{5/3}(y)\,dy \\
    &\frac{\Im Z}{Z_0} = \frac{-\gamma f}{f_\mathrm{crit}}\left[
    \int_0^1 e^{-\frac{f}{f_\mathrm{crit}}y} \right.
    \left(\frac{-2\tilde{y}}{4y\tilde{y}}\right.    +\left. \frac{\left(i\,y+\tilde{y}\right)^{5/3}+\left(i\,y+\tilde{y}\right)^{-5/3}}{4y\tilde{y}}\right)dy \\
    & \phantom{\frac{\Im Z}{Z_0} =\frac{-\gamma f}{f_\mathrm{crit}}x} - \int_1^\infty e^{-\frac{f}{f_\mathrm{crit}}y}
    \left(\frac{8i\tilde{y}}{8yi\tilde{y}}\right.
    -\left.\left.\frac{\left(y+i\tilde{y}\right)^{5/3}-\left(y+i\tilde{y}\right)^{-5/3}}{8yi\tilde{y}}\right)\,dy\right] \\
    &\textrm{with   } f_\mathrm{crit} \equiv \frac{3}{4\pi} \gamma^3 \frac{c}{\rho} \quad \textrm{and} \quad \tilde{y} \equiv \sqrt{1-y^2} \, ,
    \label{eq:SR-imp}
\end{align}
where $\gamma$ is the relativistic Lorentz factor of the beam and $K_{5/3}$ is the modified Bessel function.

%% file: text/rf_effects.tex
\section{RF effects}
\label{sec:rf-effects}

\subsection{Cavity-transmitter-beam interaction}

    Superconducting RF cavities can, close to resonance, be modelled as a simple resonator,
\begin{equation}
    Z(f) = \frac{R_s}{1 + j Q \left(\frac{\omega}{\omega_r}-\frac{\omega_r}{\omega}\right)} \, .
    \label{eq:SC-imp}
\end{equation}
The full cavity-transmitter-beam interaction can be described using a circuit model~\cite{Tuckmantel2011}, in which the~cavity is represented as an RLC circuit, the beam being a current source, and the generator as feeding the~circuit through a transmission line,
\begin{equation}
    I_{\mathrm{gen}}(t) = \frac{V_{\mathrm{ant}}(t)}{2 R/Q} \left( \frac{1}{Q_L} - 2 i \frac{\Delta \omega}{\omega} \right)
    + \frac{d V_{\mathrm{ant}}(t)}{dt} \frac{1}{\omega R/Q} + \frac{1}{2} I_{\mathrm{b,rf}}(t) \, .
    \label{eq:cav-transm-beam-1}
\end{equation}
In time-discretised form, this becomes
\begin{equation}
    V_{\mathrm{ant}}^{(n)} = \frac{R}{Q} \omega T_{\mathrm{rev},(n)} \, I_{\mathrm{gen}}^{(n-1)} + \left( 1 - \frac{\omega T_s}{2 Q_L} +
    i \Delta \omega T_{\mathrm{rev},(n)} \right) V_{\mathrm{ant}}^{(n-1)} - \frac{1}{2} \frac{R}{Q} \omega T_{\mathrm{rev},(n)} \, I_{\mathrm{b,rf}}^{(n-1)} \, .
    \label{eq:cav-transm-beam-2}
\end{equation}

\subsection{Cavity tune and beam-loading compensation}

    The cavity tune is usually chosen so as to minimise the required RF power. In the absence of beam, the~cavity frequency is tuned to the centre of the resonance $\omega_r$, where the voltage is maximised and the~power minimised for a given stored energy. Once beam is present, however, beam loading modifies the~optimum working point, and a compensation scheme is required.

    In the half-detuning scheme~\cite{Boussard1991}, the cavity is detuned so as to minimise the average generator power while keeping the antenna voltage vector constant,
    \begin{equation}
        P_{\mathrm{gen}}  = \frac{1}{8}R/Q\, Q_L \left( \frac{V_\mathrm{ant}}{R/Q} \frac{1}{Q_L} +  \Re(I_{b,\mathrm{rf}}) \right)^2 +
        \frac{1}{8}R/Q\, Q_L \left( -2\frac{V_\mathrm{ant}}{R/Q}\frac{\Delta \omega}{\omega_r}  + \Im(I_{b,\mathrm{rf}}) \right)^2 .
        \label{eq:half-detuning-1}
    \end{equation}
    For high-energy machines, where the beam current is approximately perpendicular to the RF voltage vector, $\Re(I_{b,\mathrm{rf}}) \approx 0$, and the expression simplifies to
    \begin{equation}
        P_{\mathrm{gen}}  = \frac{1}{8}\frac{V_\mathrm{ant}^2}{R/Q\, Q_L}+
        \frac{1}{32} R/Q\, Q_L I_{b,\mathrm{rf}}^2 = \frac{1}{8}V_\mathrm{ant} I_{b,\mathrm{rf}} \, ,
        \label{eq:half-detuning-2}
    \end{equation}
    with the optimum detuning frequency given by
    \begin{equation}
        \Delta \omega_\mathrm{HD} = \frac{1}{4} R/Q\, \frac{I_{b,\mathrm{rf}}}{V_\mathrm{ant}} \omega_r \, .
        \label{eq:half-detuning-3}
    \end{equation}
    
    The full-detuning scheme~\cite{Baudrenghien2012,Mastoridis2017-2} instead keeps only the antenna voltage \emph{amplitude} constant, allowing its phase to slip,
    \begin{equation}
        P_{\mathrm{gen}} = \frac{1}{8}\frac{V_\mathrm{ant}^2}{R/Q\, Q_L} + \frac{1}{2}R/Q\, Q_L \left(\frac{V_\mathrm{ant}}{R/Q} \frac{\dot{\varphi}}{\omega_r }- \frac{V_\mathrm{ant}}{R/Q} \frac{\Delta\omega}{\omega_r} + \frac{1}{2} \Im{( e^{-j\varphi}I_{b,\mathrm{rf}})}  \right)^2 ,
        \label{eq:full-detuning}
    \end{equation}
    again assuming $\Re(I_{b,\mathrm{rf}}) \approx 0$. This scheme has the advantage of compensating the bunch-by-bunch variation of the beam current along the ring through the phase-slip term $\dot\varphi$.

\subsection{RF phase and frequency modulation}

    Beam dynamics studies frequently require the RF phase and/or frequency to be modulated at a given turn $n$. Since a change in RF phase corresponds to a change in RF frequency and vice versa,
    \begin{equation}
        \Delta \varphi_{\mathrm{rf}} = 2 \pi h \frac{\omega_{\mathrm{rf}}}{\omega_{\mathrm{rf},d}} = 2 \pi h \frac{\omega_{\mathrm{rf},d} + \Delta \omega_{\mathrm{rf}}}{\omega_{\mathrm{rf},d}} \, ,
        \label{eq:phase-freq-modulation}
    \end{equation}
    phase and frequency modulation are, in principle, interchangeable. In practice, one or the other is usually preferred for a given hardware implementation. When a phase modulation $\Delta \varphi_\mathrm{rf}$ is applied, the RF frequency must be corrected by the equivalent frequency change with respect to the reference clock,
    \begin{equation}
        \Delta\omega_{\mathrm{rf},(n)} = \frac{d\Delta\varphi_{\mathrm{rf},(n)}}{dt} = \delta\Delta\varphi_{\mathrm{rf},(n)} \frac{\omega_{\mathrm{rf},d,(n)}}{2\pi h} \, ,
        \label{eq:phase-modulation}
    \end{equation}
    and, conversely, when a frequency modulation $\Delta \omega_\mathrm{rf}$ is applied, the RF phase must be corrected as
    \begin{equation}
        \Delta \varphi_{\mathrm{rf}} = \frac{2 \pi h \Delta \omega_{\mathrm{rf}}}{\omega_{\mathrm{rf,d}}} \, .
        \label{eq:freq-modulation}
    \end{equation}

\subsubsection{Accumulated phase error}

    With respect to the reference frame introduced in Section~\ref{sec:longitudinal-tracking}, an RF phase error accumulates whenever the~RF frequency is decoupled from the revolution frequency, $\omega_\mathrm{rf} \neq h \omega_\mathrm{rev}$, or whenever an explicit offset, modulation, or noise term is added to the RF phase. The phase at a given moment in time is then
\begin{align}
    \varphi_{\mathrm{rf},k}(t_{(n)}) &= \int_{0}^{t_{(n)}} \omega_{\mathrm{rf}}(\tau) \, d\tau  + \varphi_{\mathrm{offset},k,(n)} + \varphi_{\mathrm{noise},k,(n)} \\
    &= \sum_{i=1}^{n} \omega_{\mathrm{rf},k,(i)} T_{\mathrm{rev},(i)} + (t_{(n)}-t_{\mathrm{ref},(n)}) \omega_{\mathrm{rf},k,(n)} + \varphi_{\mathrm{offset},k,(n)} + \varphi_{\mathrm{noise},k,(n)} \, ,
    \label{eq:RF-phase-1}
\end{align}
which, after subtracting multiples of $2\pi$, becomes
\begin{equation}
    \varphi_{\mathrm{rf},k}(\Delta t_{(n)}) = \sum_{i=1}^{n} \frac{\omega_{\mathrm{rf},k,(i)} - h_{k,(i)} \omega_{\mathrm{rev},(i)}}{h_{k,(i)} \omega_{\mathrm{rev},(i)}} 2 \pi h_{k,(i)} + \omega_{\mathrm{rf},k,(n)} \Delta t_{(n)} + \phi_{\mathrm{offset},k,(n)} + \phi_{\mathrm{noise},k,(n)} \, .
    \label{eq:RF-phase-2}
\end{equation}
By definition, the RF phase is evaluated at $\Delta t = 0$, so that
\begin{equation}
    \varphi_{\mathrm{rf},k} = \sum_{i=1}^{n} \frac{\omega_{\mathrm{rf},k,(i)} - h_{k,(i)} \omega_{\mathrm{rev},(i)}}{h_{k,(i)} \omega_{\mathrm{rev},(i)}} 2 \pi h_{k,(i)} + \phi_{\mathrm{offset},k,(n)} + \phi_{\mathrm{noise},k,(n)} \, .
    \label{eq:RF-phase-3}
\end{equation}

\subsubsection{Phase modulation at a single frequency}

\begin{figure}[ht]
        \centering
        \includegraphics[width=0.48\linewidth]{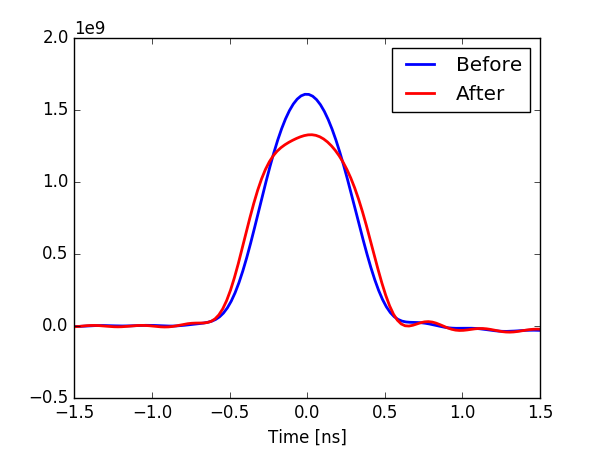}
        \includegraphics[width=0.48\linewidth]{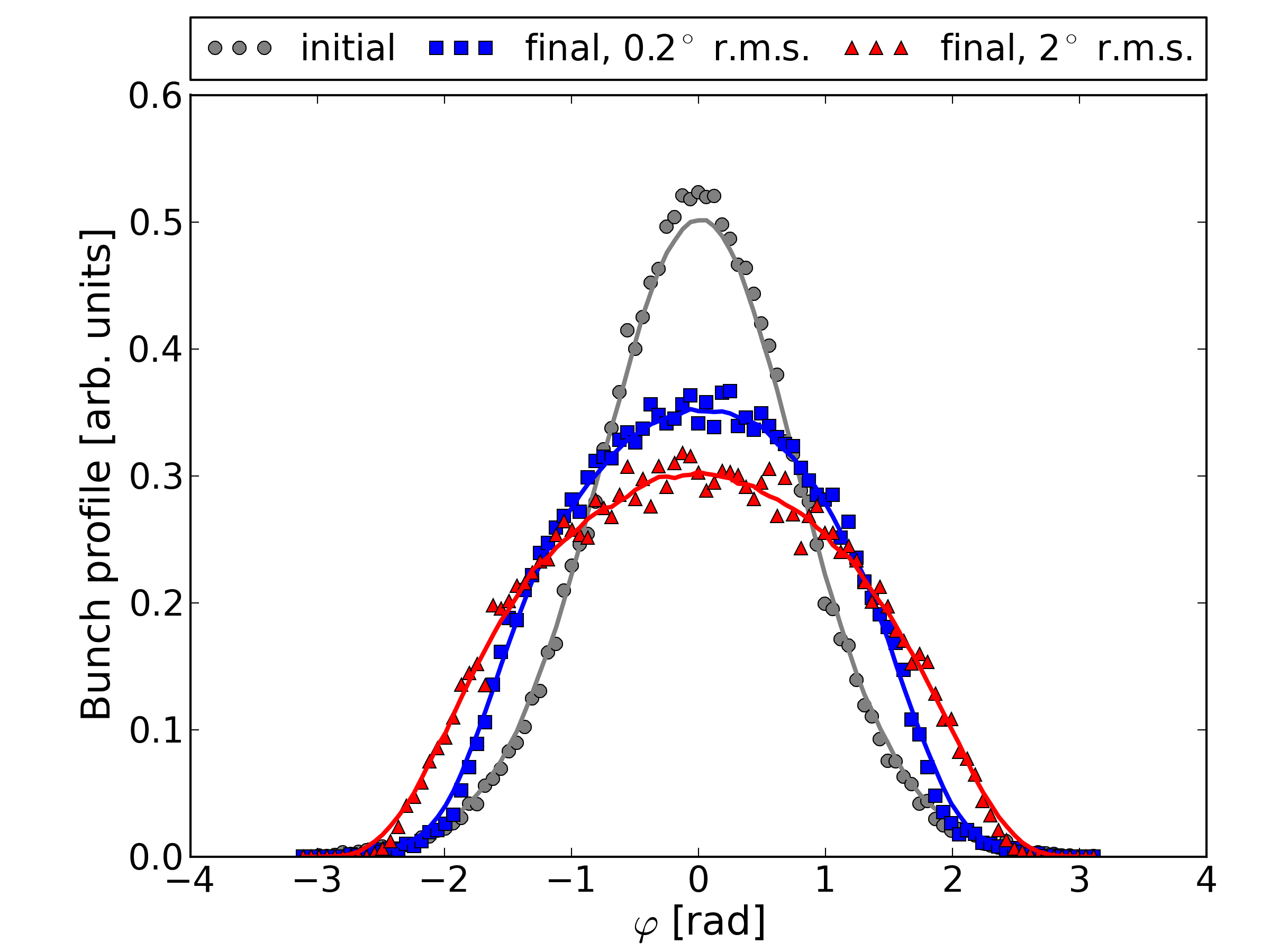}
        \caption{Bunch profile shaping using RF phase modulation and noise. Left: measured LHC bunch profile before and after RF phase modulation. Right: simulated diffusion of a bunch profile due to band-limited RF phase noise.}
        \label{fig:noise-modulation}
    \end{figure}

    A common application of RF phase modulation is resonant excitation of the bunch using a sine wave~\cite{Lee2019,Tan2012,Shaposhnikova2014}, added to the RF phase as~\cite{Albright2019}
    \begin{equation}
        \Delta\varphi_{\mathrm{rf},(n)} = A\sin\left(2\pi\sum_{k=0}^n{f_{\mathrm{mod},(k)} T_{\mathrm{rev},(k)}}\right) + \varphi_{\mathrm{off},(n)} \, .
        \label{eq:sine-modulation}
    \end{equation}
    To correctly simulate this modulation, the RF frequency has to be corrected in parallel, following Eq.~\eqref{eq:phase-modulation}. A practical example, shown in Fig.~\ref{fig:noise-modulation} (left) is bunch flattening during LHC collisions, where the~bunch length is deliberately increased to counteract the shrinkage caused by synchrotron radiation damping in a lossless manner. For this purpose, the modulation is applied close to the bunch core frequency, at $0.98\,f_{s0}$ with an amplitude of about $0.6^\circ$.

\subsubsection{RF phase noise}

    Controlled emittance blow-up can also be achieved by injecting RF phase noise into the system~\cite{Krinsky1982,Dome1985}. The RF phase can be changed turn by turn with a noise sample, $\Delta \varphi_\mathrm{rf} = \varphi_\mathrm{noise}(t_{(n)})$, or, alternatively, built up as a sum of sine-wave modulations at different frequencies. In particular, band-limited white noise can be injected to shape the bunch in a controlled way~\cite{Toyama2000} and Fig.~\ref{fig:noise-modulation} (right): the spectrum of the~noise determines which part of the bunch is affected by the diffusion, and should generally be chosen to target the bunch core, so as to avoid inducing particle losses from the tails. By construction, the average phase noise vanishes, $\langle\varphi_\mathrm{noise}(t_{(n)})\rangle = 0$, while its r.m.s.\ value is set by the spectral density $S_\varphi(f)$ of the~noise,
    \begin{equation}
        \varphi_\mathrm{noise}^{\mathrm{rms}} = \sqrt{\int S_{\varphi}(f) df} \, .
        \label{eq:noise-spectral-density}
    \end{equation}
    In general, the diffusion caused by RF phase noise has to be evaluated with full particle tracking; for some simplified cases, however, semi-analytical solvers can be used to determine the final bunch profile and length directly. For short bunches, in particular, the r.m.s.\ bunch length evolution can be obtained by applying diffusion theory~\cite{Ivanov1992},
    \begin{equation}
        \sigma(t) = \sqrt{\sigma_0^2 + \int S_{\varphi}(f) df} \, .
        \label{eq:noise-short-bunches}
    \end{equation}
    
\subsubsection{Momentum slip stacking}

    Momentum slip stacking is a technique used to reduce the bunch spacing by ``interleaving'' two bunch batches~\cite{Burnet2011,Boussard1979,Seiya2007,Ainsworth2019}. Two beams are captured with two RF systems operating at a slightly different frequency,
    \begin{equation}
        V_{\mathrm{rf}} = V_{\mathrm{rf},1}\sin(\omega_{\mathrm{rf},1}t+\varphi_{\mathrm{rf},1})+V_{\mathrm{rf},2}\sin(\omega_{\mathrm{rf},2}t-\varphi_{\mathrm{rf},2}) \, ,
        \label{eq:slip-stacking-voltage}
    \end{equation}
    so that each RF system captures one beam while perturbing the other. The small frequency difference with respect to the design frequency results in a phase error
    \begin{equation}
        \Delta \varphi_{\mathrm{rf}} = \frac{2 \pi h \Delta \omega_{\mathrm{rf}}}{\omega_{\mathrm{rf,d}}} \, ,
        \label{eq:slip-stacking-phase}
    \end{equation}
    which, at constant magnetic field, translates into a slippage, or drift, given by
    \begin{equation}
        \frac{\Delta \omega_{\mathrm{rf}}}{\omega_{\mathrm{rf,d}}} = - \eta_0 \frac{\Delta p}{p_{\mathrm{d}}} \, .
        \label{eq:slip-stacking-frequency}
    \end{equation}
    The two beams therefore slip inside the same beam pipe in opposite directions until they reach the~desired longitudinal and azimuthal position, at which point they are recaptured with a higher RF voltage at their common frequency. This technique has, for instance, been applied to slip-stack ion beams in the~SPS~\cite{Coupard2016}.

\subsection{Global control loops}

    Global control loops operate on a principle common to feedback systems in general: a beam-based measurable is compared to its design value, and the RF frequency is corrected accordingly,
    \begin{equation}
        \Delta \omega_\mathrm{TOT} \equiv \omega_\mathrm{rf} - \omega_\mathrm{rf,d} \, ,
        \label{eq:global-loops-frequency}
    \end{equation}
    where the total correction can, in general, be a sum of contributions from several individual loops.
    
    The \emph{beam phase loop} measures the beam phase with respect to the synchronous (design) phase,
    \begin{equation}
        \Delta \varphi_\mathrm{PL} = \varphi_b - \varphi_d \, ,
        \label{eq:global-loops-PL}
    \end{equation}
    and is used to reduce injection errors and to counteract undesired RF noise, thereby improving beam lifetime. The \emph{synchronisation} or \emph{frequency} \emph{loop} instead measures the RF frequency directly,
    \begin{equation}
        \Delta \omega_\mathrm{SL} = \omega_\mathrm{rf} - \omega_\mathrm{rf,d} = \omega_\mathrm{rf} - h\omega_\mathrm{rev} \, ,
    \label{eq:global-loops-SL}
    \end{equation}
    and is used at injection and extraction, and more generally to keep the RF frequency at its design value. For machines that cross transition, a \emph{radial loop} measures the radial position of the beam,
    \begin{equation}
        \frac{\Delta R_\mathrm{RL}}{R_d} = \frac{\Delta \omega_\mathrm{rf}}{\omega_\mathrm{rf,d}}\frac{\gamma^2}{\gamma_T^2 - \gamma^2} \, ,
        \label{eq:global-loops-RL}
    \end{equation}
    used to keep the beam centred while maintaining the RF frequency at its design value. Finally, the \emph{longitudinal damper} again measures the beam phase with respect to the RF phase, $\Delta \varphi_\mathrm{PL} = \varphi_b - \varphi_d$, and is used to damp phase errors bunch by bunch or batch by batch, depending on the available bandwidth, thereby counteracting coupled-bunch instabilities.
    
    In a machine with several RF systems, the measurement is typically performed at a given harmonic and the corresponding correction is applied to all RF systems. The updated RF frequency of system $k$ is
    \begin{equation}
        \Delta\omega_\mathrm{rf,k} = \frac{h_k}{h_\mathrm{meas}} \Delta \omega_\mathrm{TOT} \, ,
        \label{eq:global-loops-freqcorr}
    \end{equation}
    and the corresponding updated RF phase is
    \begin{equation}
        \Delta\varphi_\mathrm{rf,k} = 2\pi h_k \frac{\omega_\mathrm{rf,k}}{\omega_\mathrm{rf,d,k}} \, .
        \label{eq:global-loops-phasecorr}
    \end{equation}
    For each loop, the frequency correction is calculated from the corresponding measurable via a transfer function specific to that loop; a simple example is a pure gain, as used for the LHC beam phase loop,
    \begin{equation}
        \Delta \omega_\mathrm{PL} = - g_\mathrm{PL}\Delta \varphi_\mathrm{PL} \, ,
        \label{eq:global-loops-LHCPL}
    \end{equation}
    whereas the PSB beam phase loop instead uses a transfer function given in the $z$-domain,
    \begin{equation}
        H(z) = g \frac{b_0 + b_1 z^{-1}}{1+ a_1 z^{-1}} \, ,
        \label{eq:global-loops-PSBPL}
    \end{equation}
    which translates into the time domain as
    \begin{equation}
        \Delta \omega_\mathrm{PL,(n+1)} = - a_1 \omega_\mathrm{PL,(n)} + g (b_0 \Delta \varphi_\mathrm{PL,(n+1)} + b_1 \Delta \varphi_\mathrm{PL,(n)}) \, .
        \label{eq:global-loops-PSBPL2}
    \end{equation}
    
    The RF phase itself is measured by extracting the RF component from the beam profile, i.e.\ by convolving the beam profile with the RF wave and averaging over the whole beam,
    \begin{align}
        f(t) &= \int_{\lambda_{\mathrm{min}}}^{\lambda_{\mathrm{max}}}
             { \cos(\omega_{\mathrm{rf}} (t-\tau) -
             \varphi_{\mathrm{rf}}) \lambda(\tau) d\tau} \\
             &= \cos(\omega_{\mathrm{rf}} t)
             \int_{\lambda_{\mathrm{min}}}^{\lambda_{\mathrm{max}}}
             { \cos(\omega_{\mathrm{rf}} \tau +
             \varphi_{\mathrm{rf}}) \lambda(\tau) d\tau}
             +  \sin(\omega_{\mathrm{rf}} t)
             \int_{\lambda_{\mathrm{min}}}^{\lambda_{\mathrm{max}}}
             { \sin(\omega_{\mathrm{rf}} \tau +
             \varphi_{\mathrm{rf}}) \lambda(\tau) d\tau} \, ,
        \label{eq:global-loops-convolution}
    \end{align}
    from which the beam phase is determined by the ratio of the sine and cosine components,
    \begin{equation}
        \varphi_b \equiv \arctan \left(
             \frac{\int_{\lambda_{\mathrm{min}}}^{\lambda_{\mathrm{max}}}
             { \sin(\omega_{\mathrm{rf}} \tau +
             \varphi_{\mathrm{rf}}) \lambda(\tau) d\tau}}
             {\int_{\lambda_{\mathrm{min}}}^{\lambda_{\mathrm{max}}}
             { \cos(\omega_{\mathrm{rf}} \tau +
             \varphi_{\mathrm{rf}}) \lambda(\tau) d\tau}} \right) \, .
        \label{eq:global-loops-beamphase}
    \end{equation}
    
    The exact algorithms used are machine-dependent; a concrete example is the LHC, where the~beam phase loop (PL) and synchronisation loop (SL) work together~\cite{Baudrenghien2008}. There, the beam phase loop correction is $\Delta \omega_{\mathrm{PL}} = - g_{\mathrm{PL}}\Delta \varphi_{\mathrm{PL}}$, with a reaction time of 5 turns, i.e.\ $g_\mathrm{PL} = 1/(5 T_\mathrm{rev})$, while the~synchronisation loop correction is $\Delta \omega_{\mathrm{SL}} = - g_{\mathrm{SL}}(y + a \, \Delta \varphi_{\mathrm{rf}})$, with a reaction time of 50 turns, i.e.\ $g_\mathrm{SL} = 1/(50 T_\mathrm{rev})$. Here $y$ is a recursive function,
    \begin{equation}
        y_{(n+1)} = (1 - \tau) y_{(n)} + (1 - a) \tau \Delta \varphi_{\mathrm{rf}} \, ,
        \label{eq:LHC-SL-1}
    \end{equation}
    with $y_{(0)} = 0$, and $\tau$ and $a$ are functions of the synchrotron frequency, 
    \begin{align}
        a (\omega_s) &\equiv 5.25 - \frac{\omega_s}{2\pi \, 20~\text{Hz}} \, , \\
        \tau(Q_s) &\equiv 2 \pi Q_s \sqrt{ \frac{a}{1 + \frac{g_{\mathrm{PL}}}{g_{\mathrm{SL}}} \sqrt{\frac{1 + 1/a}{1 + a}} }} \, .
        \label{eq:LHC-SL-2}
    \end{align}

    A further example of the practical use of global loops is the controlled emittance blow-up applied in the LHC during the energy ramp to counteract single-bunch loss of Landau damping. During the~process, RF phase noise is injected with both the phase and synchronisation loops active, and a bunch-length feedback regulates the r.m.s.\ amplitude of the injected noise so as to keep the bunch length (and the relative synchrotron frequency spread) constant.

\subsection{Local control loops}

    In addition to the global loops discussed above, local control loops act on the RF chain from the transmitter to the cavity, with the principal function of regulating the RF voltage amplitude and phase in the~cavity, $\overrightarrow{V}_\mathrm{ant}$, to the desired set-point value $\overrightarrow{V}_\mathrm{rf,d}$. Note that, unlike the global-loop quantities above, the voltage amplitude and phase are here arrays evaluated at many points over a single turn. In addition to this main regulation function, local loops typically also include transmitter regulation -- sending an~appropriate input gain to the amplifier chain, whose final stage can be a klystron, tetrode, or IOT, and often including a polar loop that regulates the phase and amplitude of the transmitter directly -- as well as clamping or protection loops that prevent the transmitter, e.g.\ a klystron, from being driven beyond saturation.

    From the beam's point of view, local loops (i) reduce the cavity impedance at the fundamental RF frequency, thereby reducing the beam- and generator-induced voltage (the exact filter design being machine-dependent), and (ii), where necessary, damp coupled-bunch instabilities via dedicated coupled-bunch feedbacks, e.g.\ by reducing the impedance side-bands at $n f_\mathrm{rev} + k f_s$ with comb filters.

\subsubsection{Example: the SPS cavity controller}

    SPS cavities are normal-conducting, travelling-wave cavities, and are tuned to a fixed RF frequency. The~antenna voltage is a combination of the beam- and generator-induced voltages~\cite{Dome1977},
    \begin{equation}
        V_\mathrm{ant} = I_\mathrm{b} Z_\mathrm{b} + I_\mathrm{gen} Z_\mathrm{gen} \, ,
        \label{eq:local-loops-SPS-ant}
    \end{equation}
    where $I_\mathrm{b}$ and $I_\mathrm{gen}$ are the beam- and generator RF currents and $Z_\mathrm{b}$ and $Z_\mathrm{gen}$ are the beam- and generator impedances, respectively. The beam-induced impedance as a function of angular frequency can be expressed as~\cite{Baudrenghien2020}
    \begin{align}
        Z_\mathrm{b} (\omega) &=
        \frac{\rho l^2}{8} \left[
        \left( \frac{\sin \left(\frac{1}{2}\tau(\omega - \omega_r)\right)}{\frac{1}{2}\tau(\omega - \omega_r)} \right)^2
        - 2i \frac{\tau(\omega - \omega_r) - \sin(\tau(\omega - \omega_r))}{(\tau(\omega - \omega_r))^2}
            \right] + \\ \nonumber
        & +
        \frac{\rho l^2}{8} \left[
         \left( \frac{\sin \left(\frac{1}{2}\tau(\omega + \omega_r)\right)}{\frac{1}{2}\tau(\omega + \omega_r)} \right)^2
        + 2i \frac{\tau(\omega + \omega_r) - \sin(\tau(\omega + \omega_r))}{(\tau(\omega + \omega_r))^2}
        \right] \, ,
        \label{eq:local-loops-SPS-Zbeam}
    \end{align}
    where $l$ is the length and $\rho$ the shunt impedance per unit length of the accelerating structure. Similarly, the generator-induced impedance can be written as~\cite{Baudrenghien2020}
    \begin{equation}
        Z_\mathrm{gen} (\omega) =
        l \sqrt{\frac{\rho Z_0}{2}}
        \left[ \frac{\sin \left(\frac{1}{2}\tau(\omega - \omega_r)\right)}{\frac{1}{2}\tau(\omega - \omega_r)}  +
        \frac{\sin \left(\frac{1}{2}\tau(\omega + \omega_r)\right)}{\frac{1}{2}\tau(\omega + \omega_r)} \right] \, ,
        \label{eq:local-loops-SPS-Zgen}
    \end{equation}
    where $Z_0$ is the vacuum impedance.
    
\begin{figure}
        \centering
        \includegraphics[width=0.65\linewidth]{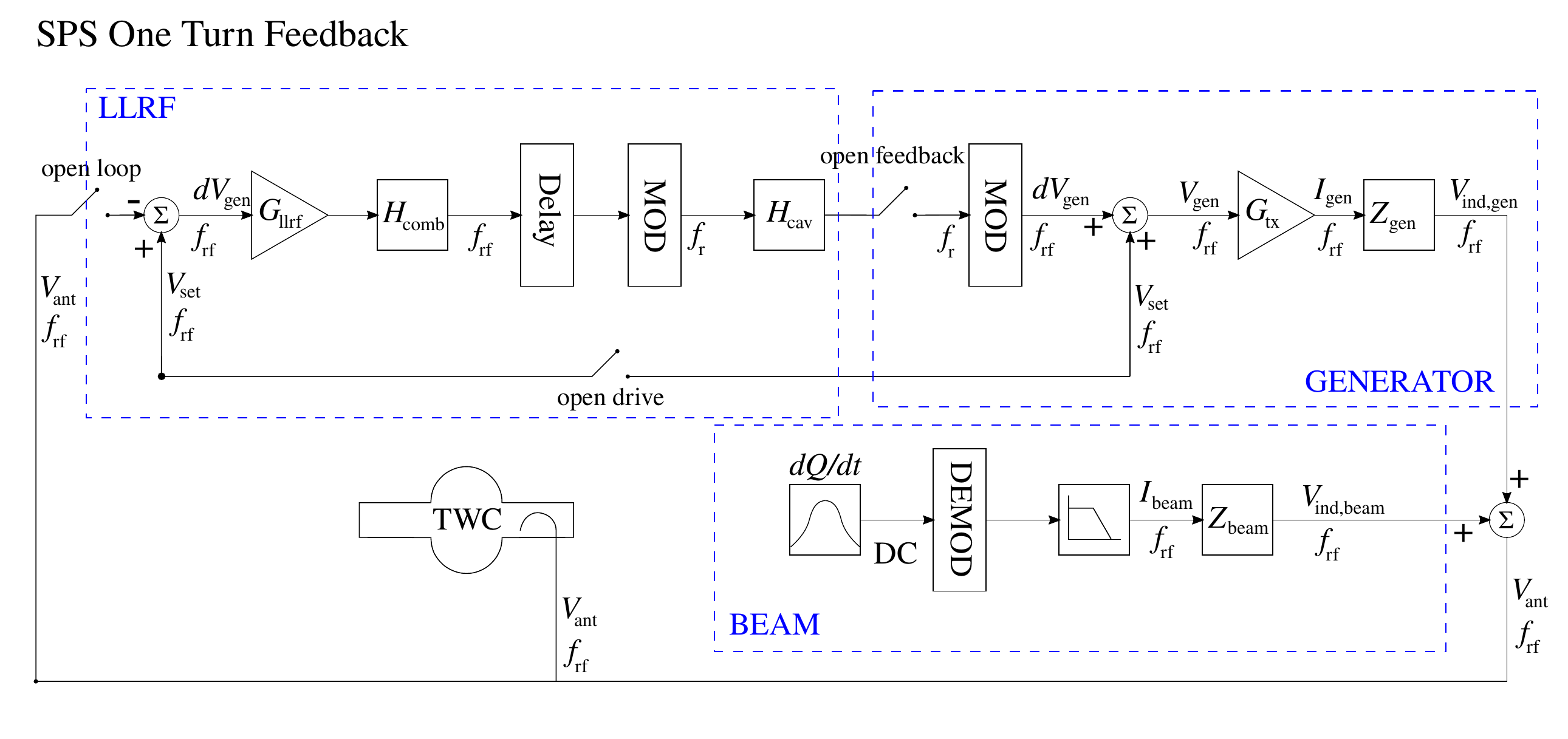}
        \caption{Model of the SPS cavity controller after~\cite{Baudrenghien2020}. The sampling time of the system is $T_s =$25~ns, corresponding to 4620 points/turn.}
        \label{fig:SPS-cavity-controller}
    \end{figure}

    As the cavities are driven at a fixed frequency $f_r$, which in general differs from the beam-synchronous frequency $f_\mathrm{rf}$, the time-domain model used in a tracking code has to include a forward- and back-modulation between these two frequencies, see Fig.~\ref{fig:SPS-cavity-controller}. The cavity response is then evaluated at the cavity frequency $f_r$, while the remaining responses are evaluated at the beam-synchronous frequency.
    
\subsubsection{Example: the LHC cavity controller}

\begin{figure}[htb!]
        \centering
        \includegraphics[width=0.65\linewidth]{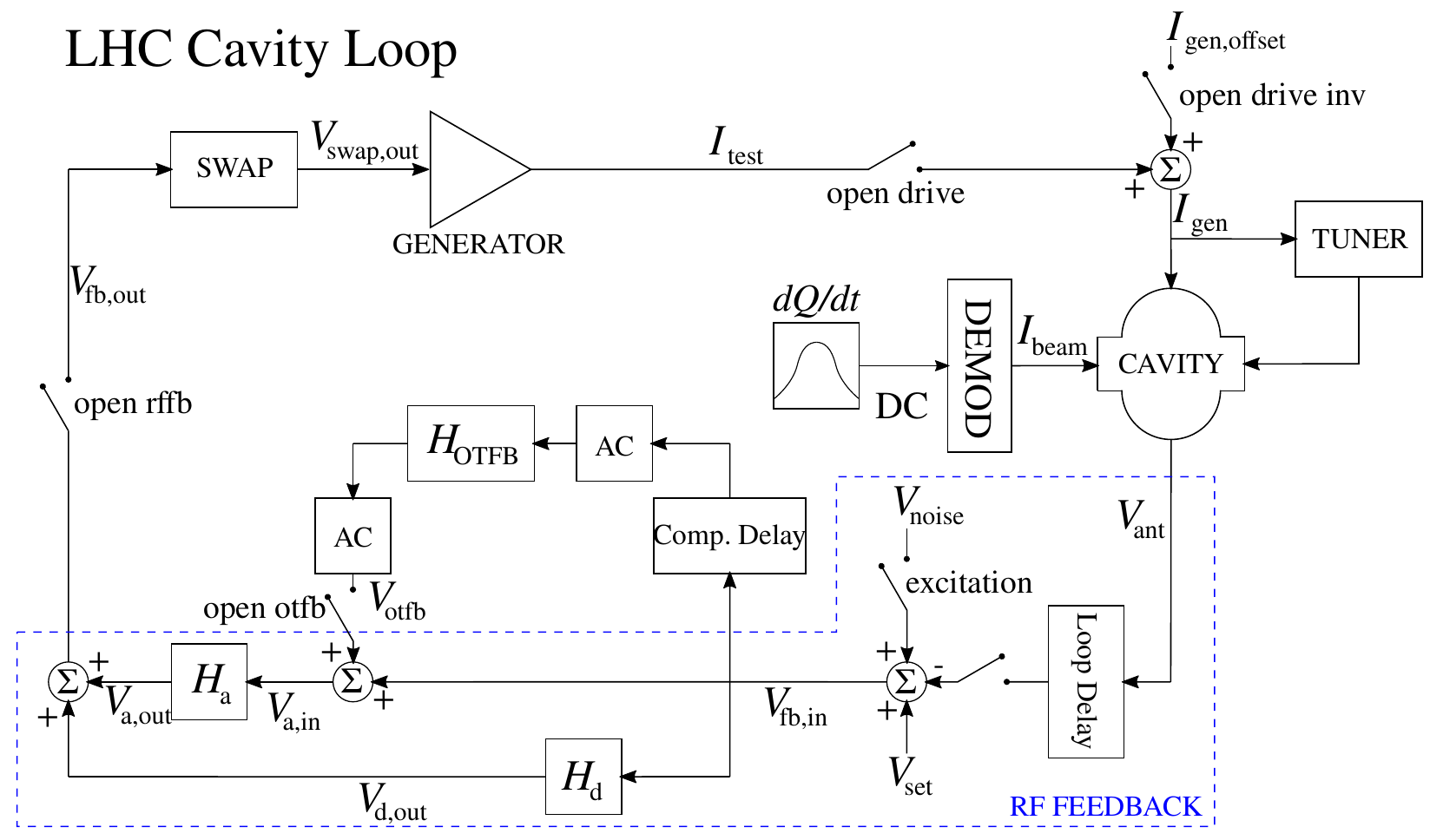}
        \caption{Model of the LHC cavity controller after~\cite{Holma2007}. The sampling time of the system is $T_s =$25~ns, corresponding to 3564 points/turn.}
        \label{fig:LHC-cavity-controller}
    \end{figure}

    In the LHC, the beam is accelerated by tuneable superconducting cavities, each fed by a klystron providing up to 300~kW of RF power. The antenna voltage at each turn and each bucket can be calculated from the cavity-transmitter-beam model of Eq.~\eqref{eq:cav-transm-beam-2}. The schematics of the LHC cavity loop are shown in Fig.~\ref{fig:LHC-cavity-controller}.
    
    The direct RF feedback is composed of an analog high-pass branch, with gain $G_a$ and delay $\tau_a$, discretised in the time domain as
    \begin{equation}
        y^{(n)} = \left[ 1 - \frac{T_s}{\tau_a} \right] \, y^{(n-1)} + G_a(x^{(n)} - x^{(n-1)}) \, ,
        \label{eq:local-loops-LHC-analogFB}
    \end{equation}
    and a digital low-pass branch, with gain $G_d$ and delay $\tau_d$, expressed as
    \begin{equation}
        y^{(n)} = \left[ 1 - \frac{T_s}{\tau_d} \right] \, y^{(n-1)} + G_a G_d e^{i \Delta \varphi_{\mathrm{ad}}}
        \frac{T_s}{\tau_d} \, x^{(n-1)} \, .
        \label{eq:local-loops-LHC-digitalFB}
    \end{equation}
    A one-turn delay feedback, implemented as a comb filter, boosts the gain of the analog branch,
    \begin{equation}
        y^{(n)} = \alpha y^{(n - N)} + G_o (1 - \alpha) x^{(n - N)} \, ,
        \label{eq:local-loops-LHC-OTFB}
    \end{equation}
    where $G_o$ is the gain, $\alpha$ a scaling factor, and $N$ the number of samples per turn; cavity tuning and klystron clamping loops complete the model.
    
\subsection{Coupling global and local control loops}

    Coupling global and local control loops poses a specific numerical challenge: local control loops typically sample one turn with a fixed integer number of samples, $T_s = T_\mathrm{rev}/N$, whereas global control loops act on the RF frequency and thereby decouple it from the revolution frequency, $\omega_\mathrm{rf} \neq h \omega_\mathrm{rev}$. As a result, the~two sets of loops are naturally defined on distinct sampling grids.

    A solution that restores a continuous sampling scheme is to modulate the relevant signals up to $N\omega_\mathrm{rev}$ for the purpose of tracking the local control loops, and to demodulate them back down to $(N/h)\,\omega_\mathrm{rf}$ for tracking the beam equations of motion. This coupling of global and local loops, and its implications for the tracking of realistic cavity controller models, is an active area of development~\cite{BLonD,Karlsen2026}.

%% file: text/particle_distributions.tex
\section{Particle distributions}
\label{sec:distributions}

\subsection{Parametrisation}

\begin{figure}[htb]
      \begin{minipage}[b]{.5\linewidth}
        \centering
        \captionof{table}{Hamiltonian distribution functions commonly used to describe proton bunches~\cite{Esteban2016}.}
        \begin{tabular}{|c|c|}
            \toprule
            \textbf{\small Distribution type} & \textbf{\small Hamiltonian distribution} \\
            \midrule
            \small{Binomial} & $F(H)=F_0 \left( 1 - \frac{H}{H_0} \right)^n$\\
            \small{Parabolic amplitude} & $F(H)=F_0 \left( 1 - \frac{H}{H_0} \right)$ \\
            \small{Parabolic line density} & $F(H)=F_0 \left( 1 - \frac{H}{H_0} \right)^{1/2}$ \\
            \small{Gaussian} & $F(H)=F_0 e^{-\frac{2H}{H_0}}$ \\
            \bottomrule
        \end{tabular}
        \vspace{24pt}
        
        \label{table:distribution-types}
      \end{minipage}
      \hspace{0.04\linewidth}
      \begin{minipage}[b]{.46\linewidth}
        \centering
        \includegraphics[trim={0 0 22.5cm 0},clip,width=0.97\linewidth]{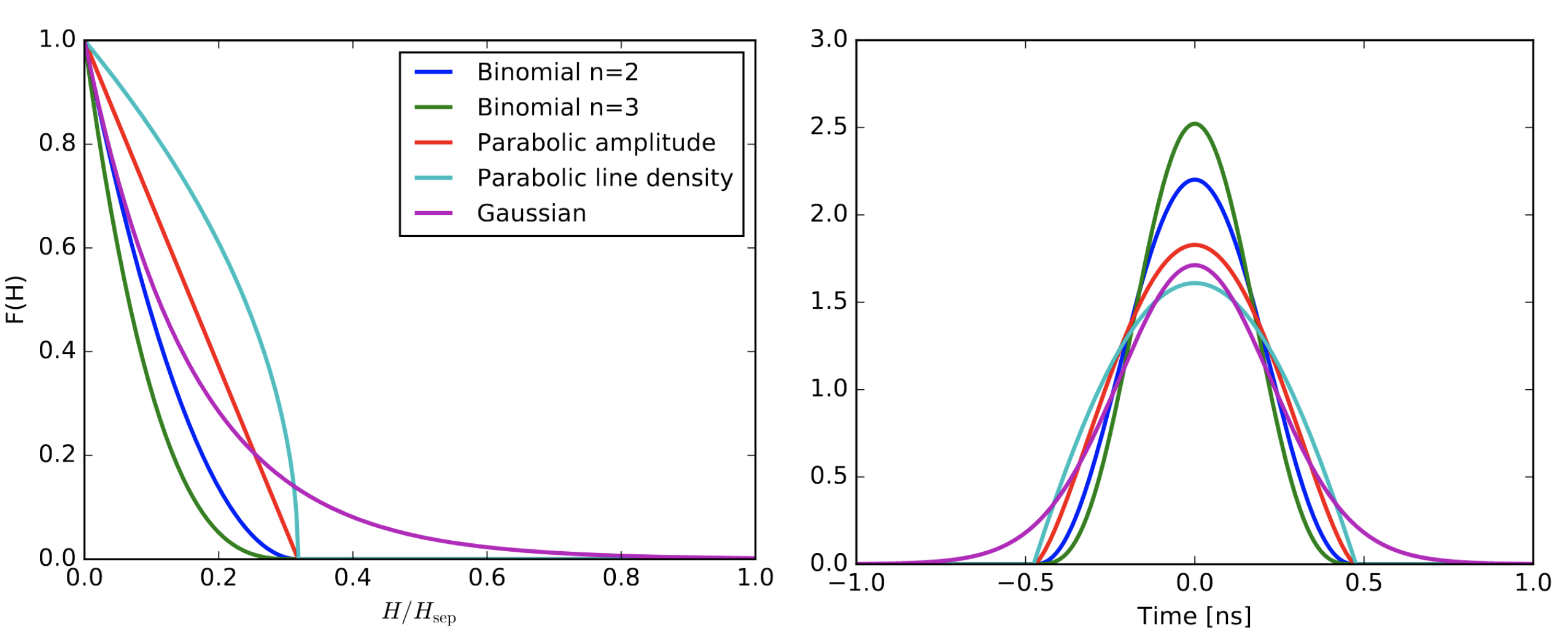}
        \captionof{figure}{Line densities corresponding to the distribution functions of Table~\ref{table:distribution-types}. Plot from~\cite{Esteban2016}.}
        \label{fig:distribution-types}
      \end{minipage}\hfill
    \end{figure}

    In a conservative system, the phase-space distribution function is conserved along a given particle trajectory, $F(\Delta t, \Delta E) = \mathrm{const.}$, as a consequence of Liouville's theorem. The distribution function is therefore itself a function of the Hamiltonian, $F(\Delta t, \Delta E) = F(H)$. Table~\ref{table:distribution-types} lists a number of Hamiltonian distributions that are particularly useful for describing (proton) bunches~\cite{Esteban2016}, together with the~corresponding line densities shown in Fig.~\ref{fig:distribution-types}: the binomial distribution, which for the special cases $n=1$ and $n=1/2$ reduce to the parabolic-amplitude and parabolic-line-density distributions, respectively, and the~Gaussian distribution, obtained in the limit $n\to\infty$ of the binomial family.
    
\subsection{Observables}
    
    A number of basic observables, common to both simulations and measurements, are used to characterise a bunch or beam: the beam profile itself, the bunch length and emittance derived from it, and, for stability studies, the dipole and quadrupole moments introduced in Section~\ref{sec:intensity-effects}.

\subsubsection{Beam profile}

    The beam profile $\lambda(\Delta t)$ is one of the most basic measurables available, although the bunch tails are typically much harder to measure accurately than the core. The Gaussian profile is characterized by its standard deviation $\sigma_t$,
    \begin{equation}
        \lambda(\Delta t) = \lambda_0 e^{-\Delta t^2/\sigma_t^2} \, ,
    \end{equation}
    where $\lambda_0$ is a normalisation constant. The binomial profile with a full bunch length $\tau_\mathrm{full}$ and a shape coefficient $\mu$ can be expressed as
    \begin{equation}
        \lambda(\Delta t) = \lambda_0 \left(1-\frac{ \sin^2{\frac{\Delta t}{2}} }{ \sin^2\textbf{}{\frac{\tau_{\mathrm{full}}}{4}} }\right)^{\mu + \frac{1}{2}} \approx \lambda_0 \left( 1- 4\left(\frac{\Delta t}{\tau_\mathrm{full}}\right)^2 \right)^{\mu+\frac{1}{2}} \, .
        \label{eq:distributions-binomial}
    \end{equation}
    where on the right-hand side a short-bunch approximation has been made.

\subsubsection{Bunch length}

    The full bunch length $\tau_\mathrm{full}$ is generally difficult to measure directly in a real machine, since it depends sensitively on how the (noisy) tails of the profile are treated. The r.m.s.\ bunch length $\sigma_t$ is, in principle, straightforward to compute, but is prone to biased results when calculated as a plain second moment of the distribution, since even a single outlying particle far from the bunch core can dominate the result; it is therefore best obtained from a Gaussian fit to the core of the profile. A practical compromise, less sensitive to the tails than the full bunch length yet more robust than a raw r.m.s.\ calculation and computationally cheaper, is scaled full width at half maximum (FWHM) bunch length,
    \begin{equation}
        \tau_{4\sigma} \equiv \frac{2}{ \sqrt{2\ln2}}\tau_\mathrm{FWHM} \, ,
        \label{eq:distribution-bunch-length}
    \end{equation}
    where the scaling factor has been chosen such that, for a Gaussian profile, the bunch length is equivalent to a $4 \sigma$ extension.
    For a binomial bunch, the full bunch length can be related to this practical definition via
    \begin{equation}
        \tau_\mathrm{full} = \frac{4}{\omega_\mathrm{rf}}\arcsin{ \left[ A \sin{ \left( \frac{\sqrt{2\ln{2}}}{8}\tau_{4\sigma} \omega_\mathrm{rf} \right)  } \right]} \approx A
    \frac{\sqrt{2\ln2}}{2} \tau_{4\sigma} \, .
        \label{eq:distribution-conversion}
    \end{equation}
    where $A \equiv \left(1-2^{-1/(\mu+\frac12)}\right)^{-1/2}$. Again, the approximation on the right-hand side is valid in the short-bunch limit.

\subsubsection{Bunch emittance}

    The bunch emittance corresponds to a given bunch-length definition, and is obtained as the phase-space area enclosed by the trajectory at that bunch length,
    \begin{equation}
        \varepsilon \equiv \oint_{\Delta t = \tau/2} \Delta E(\Delta t) \, d(\Delta t) \, .
        \label{eq:distribution-emittance}
    \end{equation}

\subsection{Beam losses}

\begin{figure}[ht]
        \centering
        \includegraphics[width=.32\textwidth]{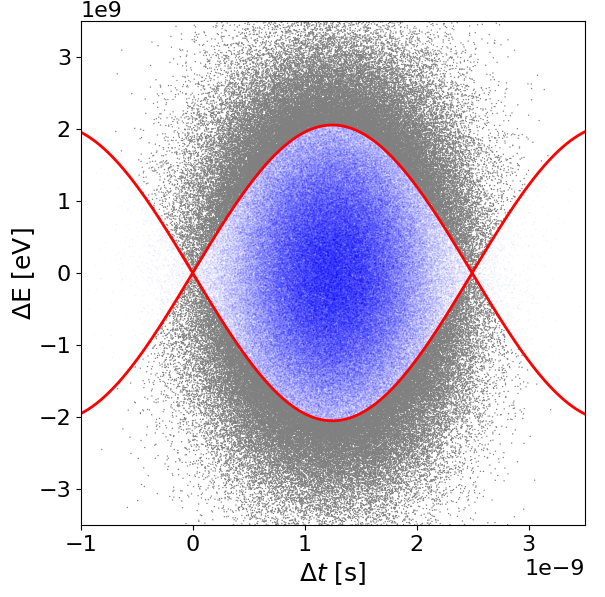}\hfill
        \includegraphics[width=.32\textwidth]{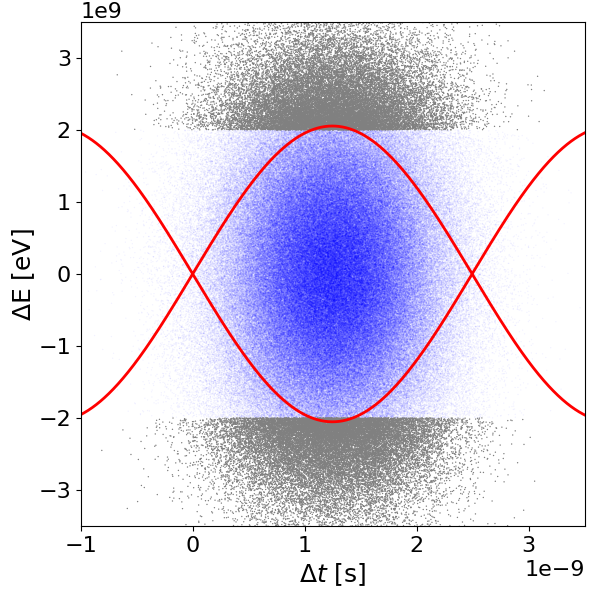}\hfill
        \includegraphics[width=.32\textwidth]{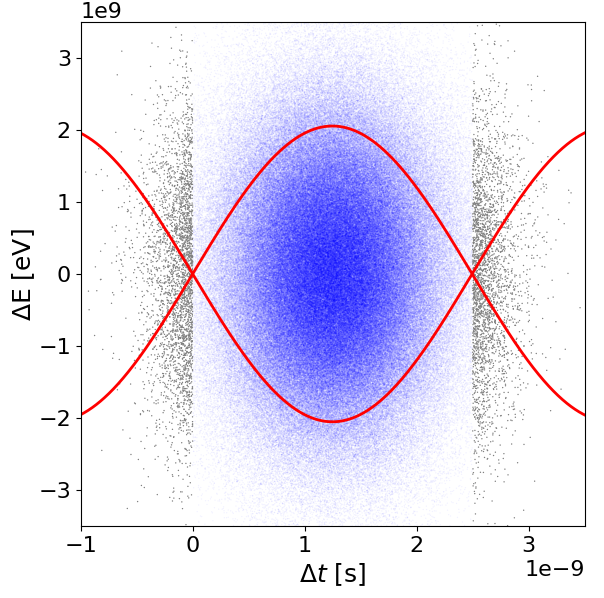}
        \caption{Common ways of defining longitudinal beam losses in simulation: by separatrix (left), by particle energy (centre), and by particle arrival time (right).}
        \label{fig:distribution-losses}
    \end{figure}

    Beam losses and debunched beam, can be defined and tracked in a simulation in several different ways, illustrated in Fig.~\ref{fig:distribution-losses}.         Depending on the simulation, it can be sometimes desired to keep debunched beam as part of the simulated population, since it can continue to affect quantities such as the induced voltage while still circulating in the machine. Physical beam losses, meaning particles that should actually not be circulating anymore, such as particles lost on a collimator, can be flagged and removed from the list of tracked particles.
    In practice, it is often convenient to use a dedicated attribute of each simulated particle to mark it as lost, since this makes it possible to still track lost particles further if needed, e.g.\ to study where in the machine aperture they are eventually absorbed.
     
    Losses can be defined by separatrix (Fig.~\ref{fig:distribution-losses} left), with or without intensity effects included. This is the strict definition of longitudinal losses, but in a real machine, it would only correspond to the~debunched population, which is hard to measure. In addition, it can be computationally heavy to evaluate, especially once induced voltage is included. Alternatively, losses can be defined by a cut in particle energy (Fig.~\ref{fig:distribution-losses} middle), corresponding to a given aperture of the machine, or by a cut in particle arrival time (Fig.~\ref{fig:distribution-losses} right), corresponding to a specific longitudinal region such as the abort gap. Both definitions are simple to evaluate and straightforward to compare with machine measurements. 
    
    A further option is to use the (integrated) beam profile directly, which is particularly useful when high-quality measured profiles are available, with an appropriate correction for the cable transfer function if needed. However, care is required when applying cuts on the profile tails to remove measurement noise, since this approach is not well suited to very low loss levels.

\subsection{Generating a bunch distribution}

    Generating a bunch distribution according to a desired analytical form can be done using the cumulative density function (CDF) of that distribution together with a random number generator (RNG). As an example, consider the generation of a bi-Gaussian bunch of $N_m$ macro-particles, without intensity effects. The seed of the RNG should first be initialised; using a fixed seed is good practice, since it allows the~same random distribution to be reproduced exactly at a later time. Two independent arrays of normally distributed random numbers with $\sigma=1$, $\mathtt{RAND}_{1,k}$ and $\mathtt{RAND}_{2,k}$ for $k = 0, 1, ..., N_{m-1}$, are then generated. The time coordinates of the bunch, for a user-specified bunch length $\sigma_t$, follow as
    \begin{equation}
        \Delta t_k = \sigma_t \mathtt{RAND}_{1,k} +
            \frac{\varphi_s - \varphi_\mathrm{rf}}{\omega_\mathrm{rf}} \, ,
        \label{eq:distribution-gaussian-time}
    \end{equation}
    while the ``matched'' energy coordinates, proportional to the bucket height, are, for a single RF system,
    \begin{equation}
        \Delta E_k = \sigma_E \mathtt{RAND}_{2,k} = \sqrt{\frac{V_\mathrm{rf} E_d \beta_d^2 }{\pi h |\eta_0|}(\cos{\varphi_b} - \cos{\varphi_s} + (\varphi_b - \varphi_s)\sin{\varphi_s}) }\,\mathtt{RAND}_{2,k} \, ,
        \label{eq:distribution-gaussian-energy}
    \end{equation}
    with $\varphi_b = \omega_\mathrm{rf} \sigma_t + \varphi_s$.
    
        The normally distributed random numbers themselves can be generated from uniform random numbers $\mathtt{RAND}_3 \in (0,1)$ via the cumulative distribution function of the normal distribution,
    \begin{equation}
        f(x) =\frac{1}{\sqrt{2\pi}}e^{-x^2/2} \rightarrow F_\mathrm{CDF} = \int_0^r r e^{-r^2/2} d r = 1 - e^{-r^2/2} \in (0,1) \text{  for  } r \in[0,\infty) \, ,
        \label{eq:distribution-gaussian-function}
    \end{equation}
    from which the required transformation follows by inverting $F_\mathrm{CDF}$,
    \begin{equation}
        x_k = \sqrt{-2 \ln{(\mathtt{RAND}_{3,k})}} \, .
        \label{eq:distribution-gaussian-input}
    \end{equation}

\subsection{Matching distributions}

\begin{figure}[ht]
        \centering
        \includegraphics[width=.48\textwidth]{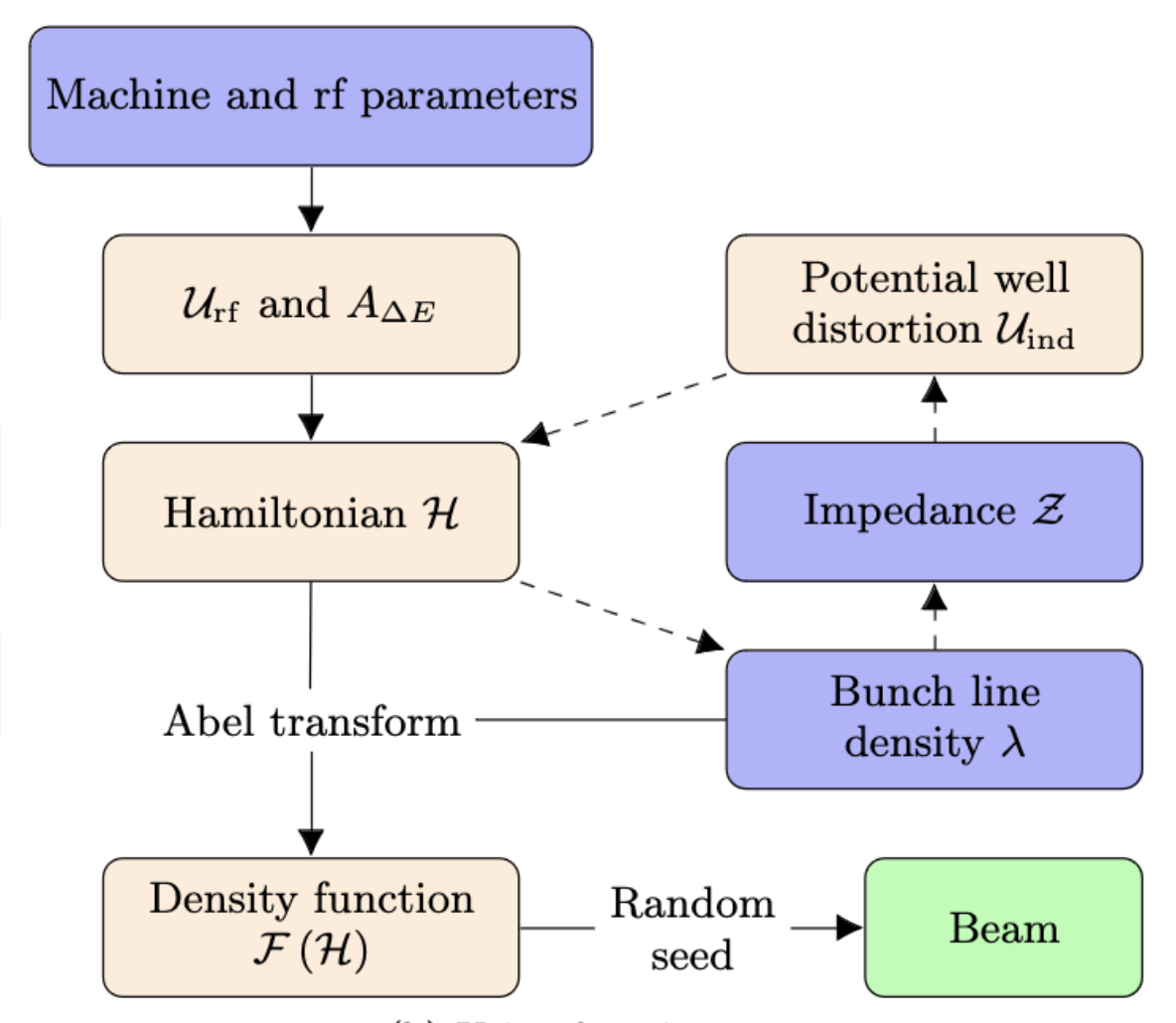}\hfill
        \includegraphics[width=.48\textwidth]{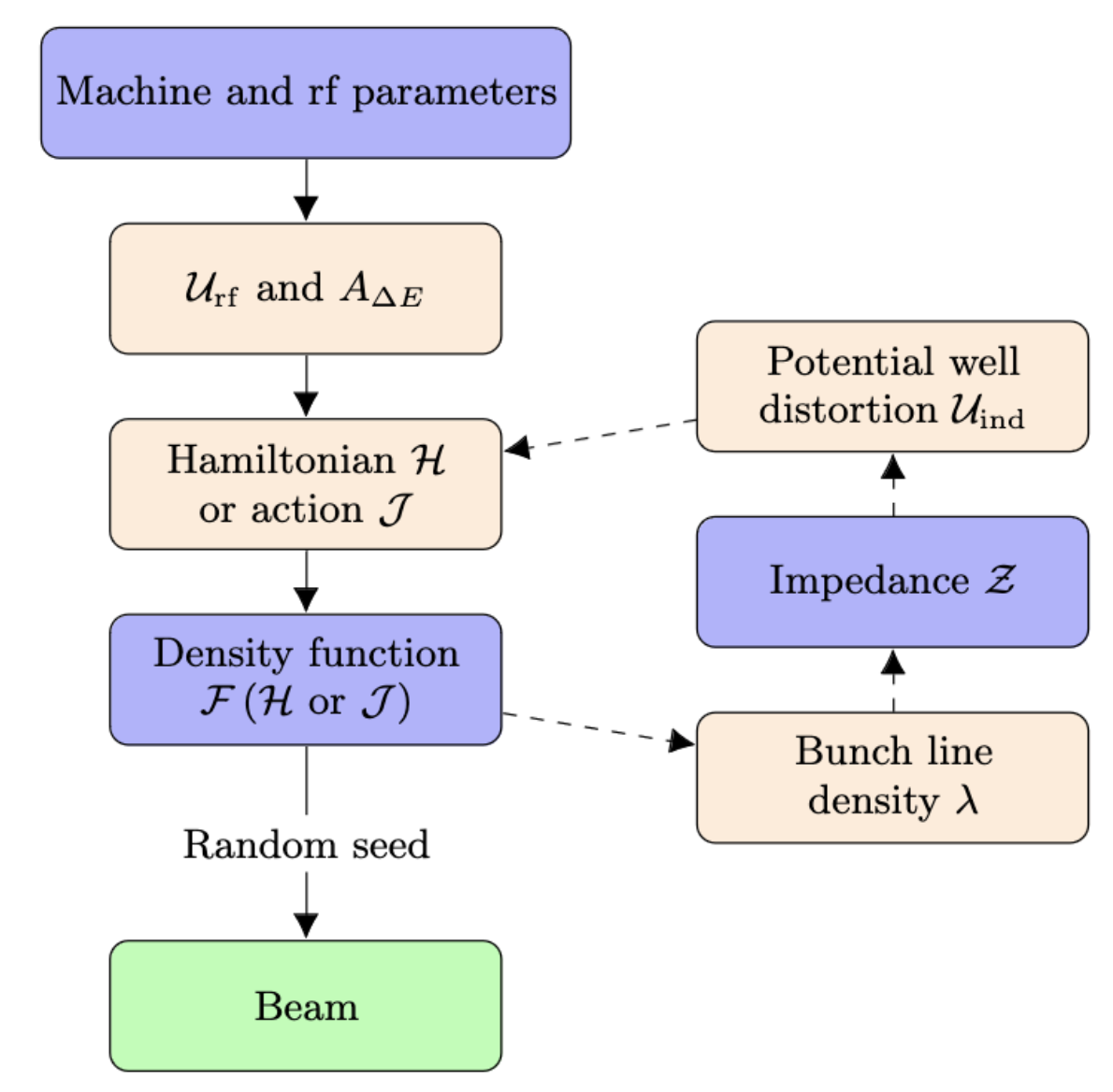}\hfill
        \caption{Flow chart of BLonD algorithms for matching a line density (left) or a density function (right).}
        \label{fig:distribution-matching-algorithm}
    \end{figure}

    Beyond the simple analytical generation described above, a bunch distribution is often required to be \emph{matched} to the actual (possibly intensity-dependent) bucket shape, with or without control loops acting on the beam. This generally relies on recursive, iterative algorithms: a function to be minimised has to be defined, which can, for instance, match a desired density function or beam profile to a prescribed accuracy. Since the induced voltage depends on the distribution while the matched distribution itself depends on the (distorted) potential well, the potential well and the distribution have to be recalculated iteratively, acting back on each other, until convergence is reached. This process also requires locating the minimum of the potential well in order to correctly place the bunch, which can become tricky in the~presence of a strong induced voltage or several RF harmonics; it is good practice to always check the~resulting matched distribution visually. Figure~\ref{fig:distribution-matching-algorithm} illustrates the corresponding algorithm as implemented in BLonD, for matching either a line density or a density function.

\subsubsection{Example: matching with intensity effects and control loops}

\begin{figure}[ht]
        \centering
        \includegraphics[width=0.6\textwidth]{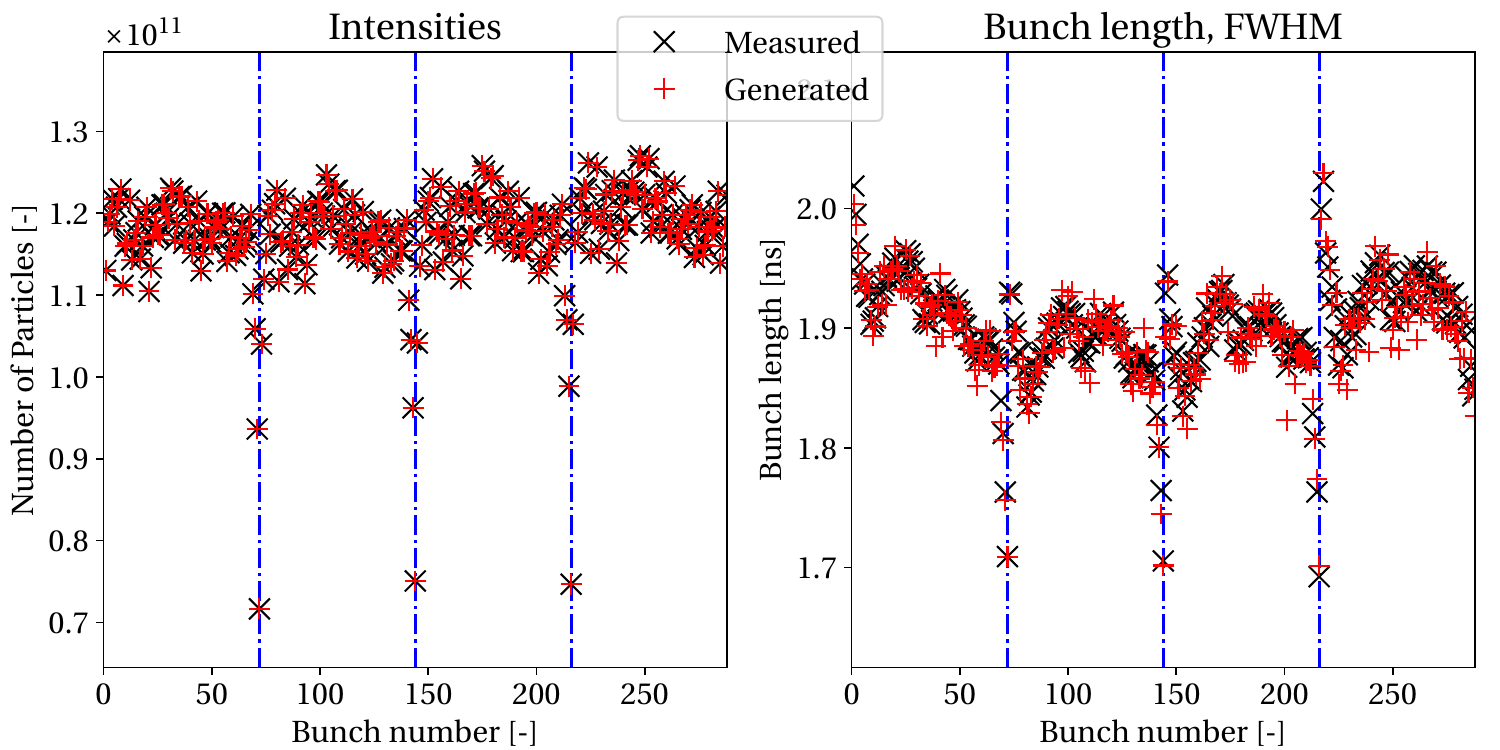}\hfill
        \caption{Matching of bunch-by-bunch intensities (left) and bunch lengths (right). The measured data (black crosses) is compared to the generated values (red plus signs), after tracking with the SPS cavity feedback included.}
        \label{fig:distribution-matching-example}
    \end{figure}

    Matching a beam distribution becomes even more involved once control loops are also acting on the~beam, since the loops themselves respond to, and modify, the distribution being matched. An illustrative example is the matching of an SPS flat-top distribution including intensity effects together with the SPS one-turn delay feedback, where a measured beam profile is used as an input, constraining the bunch-by-bunch bunch length and intensity. The approach used is to match first the beam including intensity effects alone, and then track the distribution together with the feedback, resulting in the~bunch parameters illustrated in Fig.~\ref{fig:distribution-matching-example}. In such cases, it is often useful to ramp up the feedback gain adiabatically, so as to avoid undesired emittance blow-up.

%% file: text/6d_effects.tex
\section{6D effects}
\label{sec:6d-effects}

    Some phenomena couple the longitudinal and transverse planes so strongly that they cannot be captured by a purely longitudinal (or purely transverse) model, and instead require a full six-dimensional (6D) beam dynamics treatment. Three examples of such 6D effects -- crab cavities, intra-beam scattering, and electron cloud -- are briefly discussed below.

\subsection{Crab cavities}

    Crab cavities are currently being produced for the HL-LHC project~\cite{Aberle2020}. They increase the luminosity in 
    \begin{wrapfigure}{r}{0.43\textwidth}
      \centering
      \includegraphics[width=0.4\textwidth]{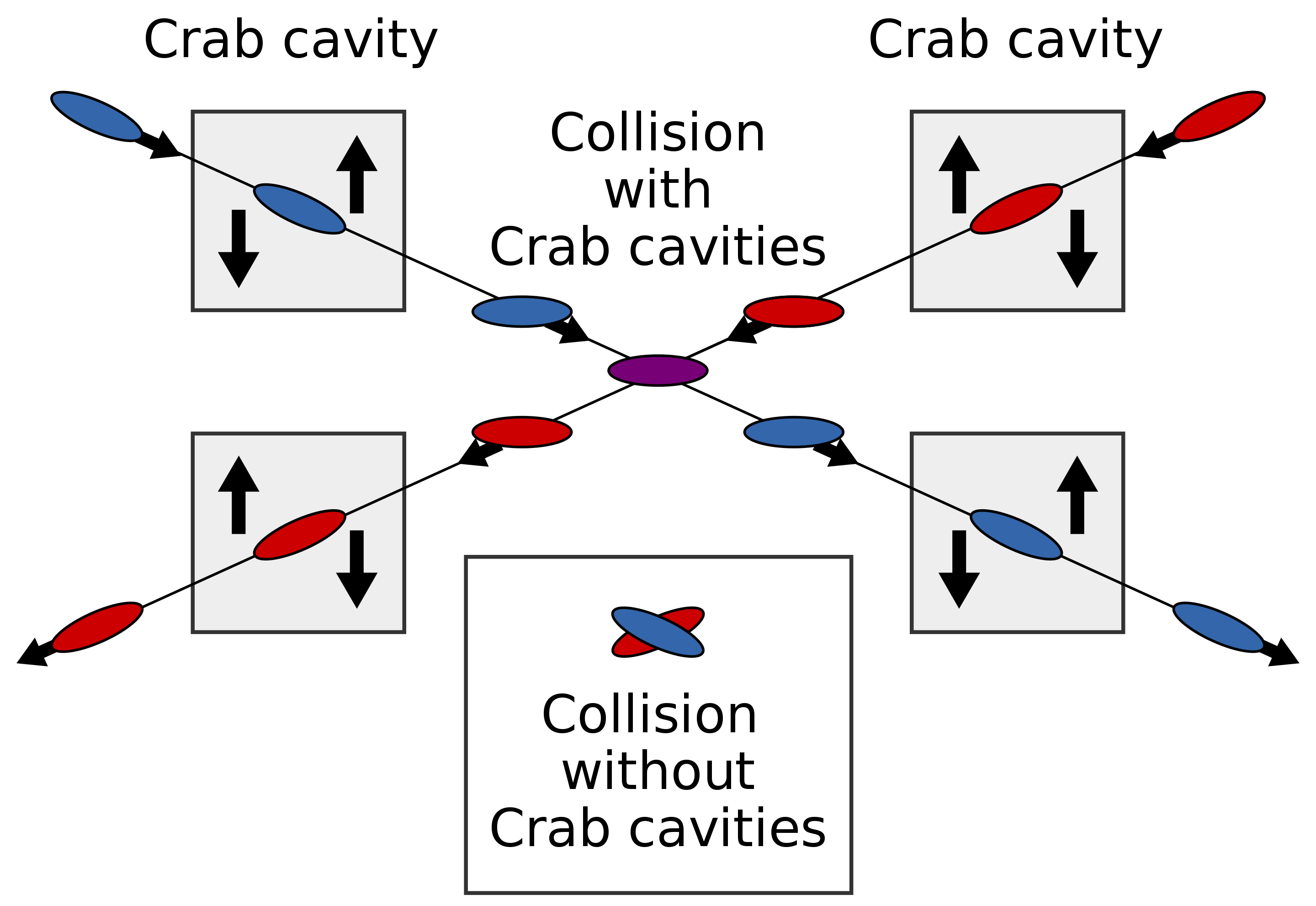}
      \caption{HL-LHC collision scheme with crab cavities~\cite{Aberle2020}. An illustration from~\cite{Alekou2020}.}
      \vspace{-24pt}
      \label{fig:6Deffects-crabbing}
    \end{wrapfigure}        
    the LHC interaction points 1 and 5 by applying a time-dependent transverse kick in the plane where a crossing angle is present between the two colliding beams, see Fig.~\ref{fig:6Deffects-crabbing}. 
    A bunch passing through a deflecting RF cavity at zero phase has its head and tail deflected in opposite directions, effectively rotating (``crabbing'') the bunch so that it collides head-on despite the crossing angle.


    Because the crabbing kick couples all three planes of motion, a full 6D beam dynamics simulator is needed to model it correctly, e.g.\ BLonD coupled to Xsuite~\cite{Xsuite}, PyHEADTAIL~\cite{Pyheadtail}, or MAD-X~\cite{MADX}. Several such numerical tools have been, and are being, developed for this purpose~\cite{Alekou2020}.

\subsection{Intra-beam scattering}

    Intra-beam scattering (IBS) describes Coulomb scattering among the particles of a beam. It comprises both single large-angle scattering events with a large change of momentum, known as the Touschek effect, and the cumulative effect of many small-angle multiple scatterings. Conceptually, it is therefore treated in a manner similar to a PIC problem, with a somewhat different treatment required in electron (short-bunch) and hadron (long-bunch) machines. Intra-beam scattering leads to an emittance blow-up in all three phase-space planes -- longitudinal, horizontal, and vertical -- simultaneously.

    The evolution of the emittances under IBS can be calculated numerically for many cases of interest~\cite{Nagaitsev2005}, and simplified models exist that are suitable for use within longitudinal beam dynamics simulations~\cite{Zampetakis2024}. Ideally, a full 6D, stochastic simulation of the scattering process would be required; in practice, however, suitable approximations exist that allow the effect to be included as an additional energy kick on the particle, without a full 6D treatment~\cite{Bruce2010}.

\subsection{Electron cloud}

\begin{figure}[ht]
    \centering
    \includegraphics[width=0.8\textwidth]{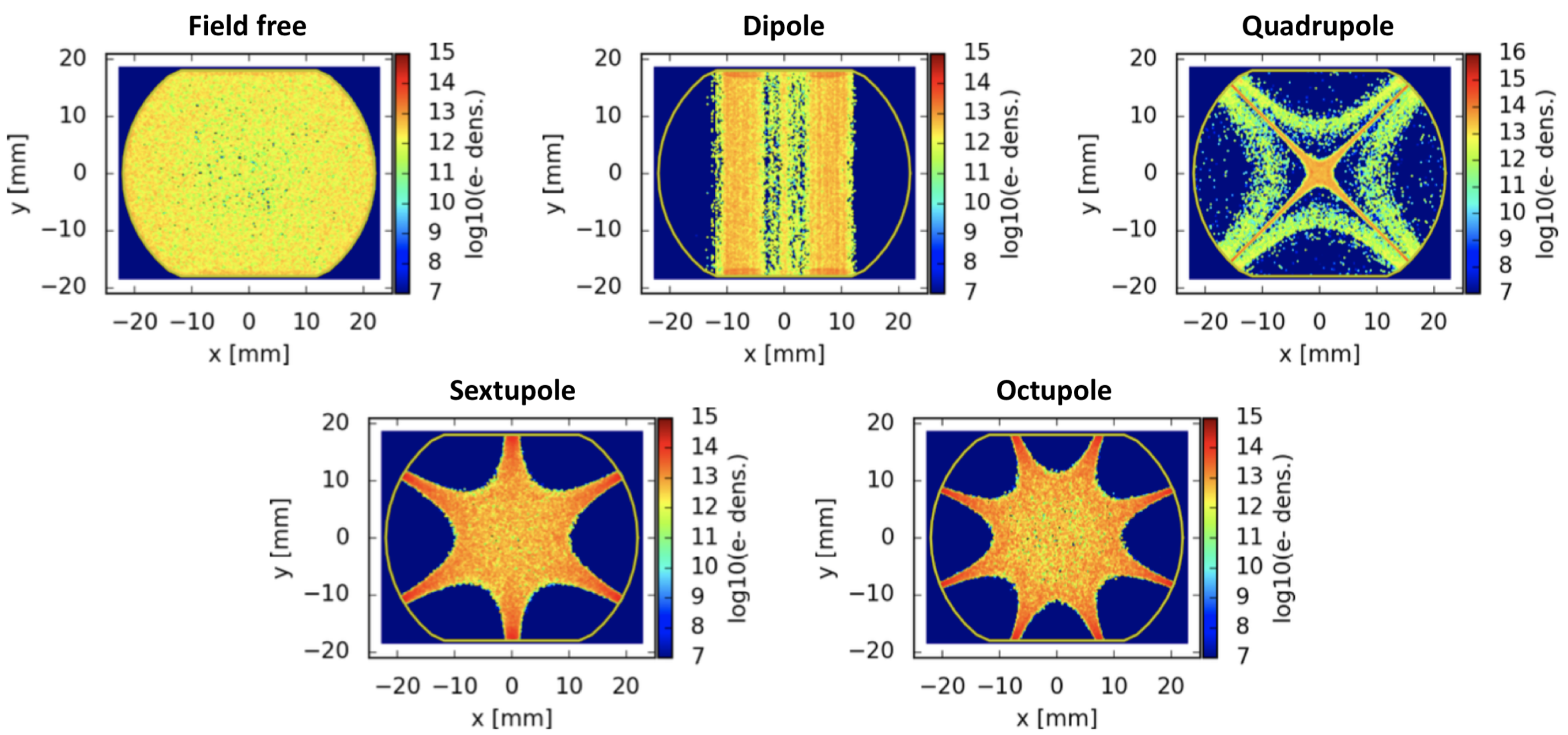}\hfill
    \caption{Electron distribution in LHC arc components with different magnetic field configurations. An illustration from~\cite{Iadarola2018}.}
    \label{fig:6Deffects-ecloud}
    \end{figure}

    In proton machines, electrons can be released from the vacuum-chamber wall through the repeated passage of the proton bunches -- e.g.\ via photoemission or secondary emission -- and subsequently multiply inside the beam pipe to form an electron cloud, see Fig.~\ref{fig:6Deffects-ecloud}. Modelling an electron cloud therefore requires tracking two particle species simultaneously: the circulating protons of the beam, and the cloud electrons, which oscillate around a location fixed by the local electromagnetic fields.

    This introduces additional modelling requirements: the surface properties of the vacuum chamber -- in particular the secondary emission yield -- need to be modelled, and a full 6D simulation of the~interaction between the two particle species is required, including their overall space charge as well as scattering and collision processes. Electron-cloud build-up is, in essence, a two-component-plasma problem, and is therefore again well suited to a PIC treatment. An example of such a simulation suite is PyECLOUD~\cite{Pyecloud}, coupled to the beam dynamics code PyHEADTAIL~\cite{Pyheadtail}, presently being ported to Xsuite~\cite{Xsuite}.

%% file: text/code_design.tex
\section{Code design, optimisation and benchmarks}
\label{sec:code-design}

\subsection{Design choices}

    Before writing any code, it is worth deciding on the overall design of the simulation suite. A first choice is between a fixed pipeline, where the sequence of tracked objects is the same for every simulation, and a flexible, modular design. A fixed pipeline can be reduced to an input file of parameters and can be heavily optimised for a given hardware platform, but requires the possible points of interaction with the~computational core -- e.g.\ when and what to output -- to be predetermined as well. A modular design, by contrast, is more easily extended to new applications, at the cost of being more difficult to optimise, and requires extra care to ensure that the computational sequence remains physically correct for every possible combination of tracked objects. A second, related choice concerns the target hardware platform: for codes intended to run on a wide range of platforms, from a laptop to multi-CPU or multi-GPU clusters, it is advisable to separate the physics code as much as possible from the platform-dependent code, so that new hardware can be supported without having to rewrite the physics core each time.

\subsection{Coding language}

    The choice of programming language typically involves a trade-off between development speed and runtime performance. High-level languages, such as Python, allow fast prototyping and produce code that is easier to read and closer to the underlying equations. They are, however, more forgiving of mistakes -- for instance, silently mixing floating-point and integer types -- which can hide subtle bugs, though this can be mitigated by interfacing with dedicated modules for CPU or GPU acceleration. Low-level languages, such as C++ or CUDA, can achieve substantially better performance if written correctly, at the cost of slower prototyping, a compilation step, and code that generally needs to be adapted explicitly to the target hardware architecture; they are, in this sense, less forgiving, but also make it clearer what is actually happening under the hood.

\begin{figure}[htb]
        \centering
        \includegraphics[width=0.95\textwidth]{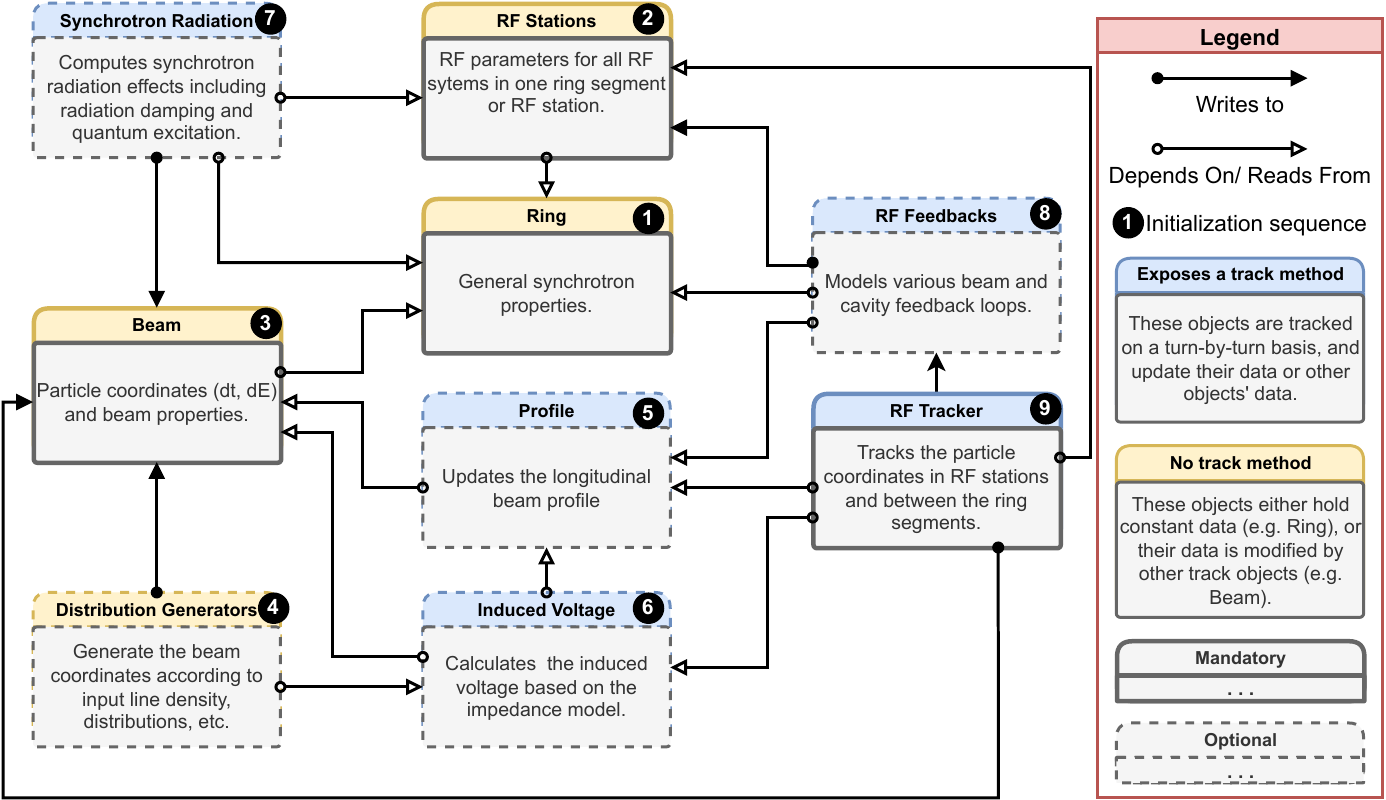}\hfill
        \caption{Schematic view of the BLonD code structure~\cite{Timko2023}. Data containers are highlighted in yellow, while trackable objects are marked in blue.}
        \label{fig:BLonD-structure}
        \vspace{-12pt}
    \end{figure}

    The BLonD code is a practical example of a hybrid approach, combining a Python front end with a C++ and CuPy computational core, see Fig.~\ref{fig:BLonD-structure}. Its structure is organised around three data containers -- \texttt{Ring}, \texttt{RFStation}, and \texttt{Beam} -- and a set of trackable objects that act on these containers turn by turn, following a modular, object-oriented design.

\subsection{Good coding practices}

    Regardless of the language chosen, a few general good practices help keep a simulation code readable and maintainable in the long run. An object-oriented design, making appropriate use of inheritance, and breaking the code down into well-defined functions, together with adequate comments and documentation, goes a long way towards avoiding ``spaghetti code''. It is similarly good practice to keep the different parts of a simulation suite clearly separated: (i) the computational core itself; (ii) configuration and input files governing a specific simulation or use case; (iii) a command-line argument parser for options that should be user-adjustable at run time; and (iv) separate submission scripts for job submission to a computing cluster. Finally, since reading, writing, and plotting data all take non-negligible time, it is worth minimising the frequency of I/O operations and thinking in advance about exactly which output is actually needed from a given simulation.

    When it comes to working in a team, a shared repository with a clearly identified production version, using proper versioning and tagging (e.g.\ with git or svn) is highly recommended, rather than informally named files such as ``my\_script\_final\_version\_2.py''. Again, combining this with adequate documentation (e.g.\ via GitHub Pages) and installation instructions, helps to keep the codebase consistent. Individual development is best organised through personal forks and feature branches, with pull requests used to propagate new features into the production version -- giving other team members time to test and react. An issue tracker is recommended, too, to record bug reports, suggestions, and planned improvements.

\subsection{Runtime optimisation}

    For computationally heavy simulations, the core or computationally dominant part of the code should first be identified through profiling, and separated from any surrounding analysis code (plotting, etc.). Then, the core can be optimised for the target hardware independently, ideally using a compiled language interfaced with the higher-level code. Beyond low-level optimisation, it is often more effective to revisit the underlying data structures and algorithms -- for instance, replacing an $N^2$ algorithm with an $N\log N$ one, as is done when using an FFT instead of a direct discrete Fourier transform (Section~\ref{sec:intensity-effects}). At the implementation level, several further tools and techniques are available: (i) vectorisation, exploiting the vector units of modern processors (requires compatible implementation of equations); (ii) multi-threading, e.g.\ via OpenMP, which can be used even on a standard office PC; (iii) hardware accelerators such as GPUs or FPGAs; and (iv) the choice between single- (32-bit) and double- (64-bit) precision floating-point numbers, which should be tested explicitly for the application at hand, since lower precision is not always acceptable. Finally, it is generally preferable to rely on well-established, heavily optimised third-party libraries (e.g.\ for FFTs) rather than re-implementing standard numerical routines.

\begin{figure}[htb!]
        \centering
        \includegraphics[width=.48\textwidth]{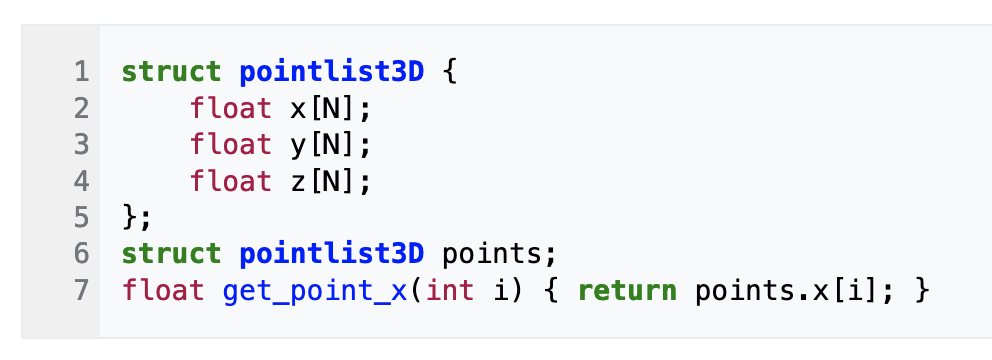}\hfill
        \includegraphics[width=.48\textwidth]{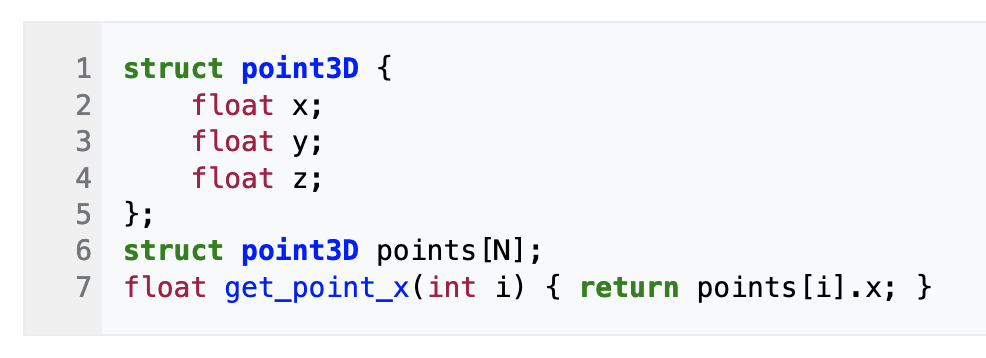}
        \caption{Example of the difference in memory access pattern between data structures. Left: structure of arrays (SoA). Right: array of structures (AoS).}
        \label{fig:data-structures}
        \end{figure}

    Memory-heavy simulations pose a related, but distinct set of challenges, which have become increasingly common as processing speed has grown faster than memory-access speed. The optimal strategy is platform-dependent, but some general good practices apply: minimising the writing of files and the exchange of data with disk, and trying to fit as much of the working data set in memory as possible, is particularly important given the large arrays -- often hundreds of millions of coordinates -- involved in beam tracking. Minimising file writes matters especially on compute clusters, where network file systems (NFS) are typically slower than local SSD or HDD storage; the choice of file format and compression level (e.g. plain text or CSV versus HDF5 or compressed archives) should be adapted to the application and hardware at hand. Where possible, tiling the input data, i.e.\ operating on a subset of the input at a~time so that it fits into a given level of cache memory, can bring a substantial speed-up -- the access to different cache levels is analogous to having to fetch information from a huge library, a given shelf, a~certain book, or having the information already accessible in your brain. Similarly, choosing data structures and access patterns that favour regular over random memory access is beneficial, e.g.\ organising the histogram of a multi-bunched beam by individual bunches rather than across the whole beam at once, as illustrated in Fig.~\ref{fig:data-structures}.
    
    For larger studies, jobs are usually submitted to a computing cluster rather than run interactively; interactive sessions remain useful for debugging, development, or small-scale studies, while batch (offline) submission is preferred for long studies or parameter scans. Cluster schedulers, such as Slurm, HTCondor, PBS, or LSF, manage job priorities, quotas, and queues based on the requested resources, memory, and expected run time, each with its own set of commands. Basic familiarity with the Linux command line is a prerequisite, and visualisation of results should generally be done off the cluster rather than on it.

\begin{figure}[ht]
        \centering
        \includegraphics[width=0.75\textwidth]{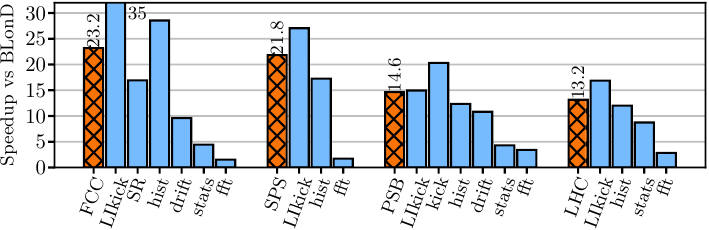}
        \includegraphics[width=0.68\textwidth]{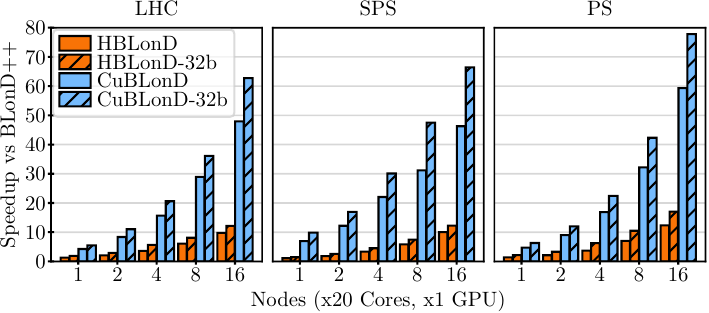}
        \caption{Top: speed-up of BLonD++ relative to BLonD for different CERN use-cases, broken down by the main computational kernels (macro-particle kick, synchrotron radiation, histogram, drift, statistics, FFTs). Bottom: speed-up of the distributed (HBLonD) and GPU-accelerated (CuBLonD) versions relative to BLonD++, as a~function of the number of computing nodes. Courtesy of K.~Iliakis~\cite{Iliakis2022}.}
        \label{fig:BLonD-speedup}
        \end{figure}

    As a concrete example of the performance gains achievable through such optimisations, the~BLonD code has evolved from a pure-Python implementation into BLonD++, with its computational core moved to C++ and time-consuming code regions optimised (including the use of highly efficient mathematical libraries), followed by HBLonD, a distributed, MPI-based version using high-performance-computing techniques such as dynamic load balancing and data- and task-level parallelism, and Cu\-BLonD, a GPU-accelerated version available in both CUDA and CuPy implementations~\cite{Iliakis2022}. Depending on the specific simulation and machine, these successive optimisation steps have brought speed-ups of roughly an order of magnitude at each stage, reducing, in one representative case, the runtime of a full simulation from about a day to around 80 minutes, see Fig.~\ref{fig:BLonD-speedup}.

\subsection{Testing, benchmarking, and comparison}

    Continuous verification of the correctness of a simulation code is essential to maintain trust in its results. There is no way to guarantee with 100\% certainty that a given simulation result is correct, but confidence increases over time if the code is tested constantly and systematically~\cite{Timko2012}. It is useful to distinguish between three related but different verification strategies: a (i) \emph{test} exercises a small, well-defined piece of the code, for which the correct input-output relation is known; a (ii) \emph{benchmark} instead exercises a global, physical problem simulated with the code, for which the expected outcome is known either from theory or from measurement; and a (iii) \emph{code-to-code comparison} is used in the absence of a~known benchmark solution, by simulating the same problem with two independent codes that may rely on different computational methods.

    Unit tests exercise individual, basic functions of the code, either against a known theoretical value or against the value obtained with an earlier version of the code -- in the latter case, correctness is of course not guaranteed, only consistency with past behaviour. Python provides a dedicated module for unit testing, which can be run on demand or integrated into the continuous-integration pipeline of a~repository.

    Code-to-code comparisons test a given physics problem with two independent simulation suites, which may rely on different underlying physics assumptions; care must be taken to ensure that the same parameters and assumptions are actually being compared. Reaching agreement between the two codes is reassuring, even though it remains possible, in principle, for both codes to share the same systematic error. As an example, the shift of the synchrotron frequency at the threshold of loss of Landau damping (Section~\ref{sec:intensity-effects}) has been compared between the macro-particle code BLonD and the semi-analytical solver MELODY~\cite{melody}, for a single broad-band resonator impedance in the LHC, and good agreement was found between the two independent approaches.

\begin{figure}[ht]
        \centering
        \includegraphics[width=0.45\textwidth]{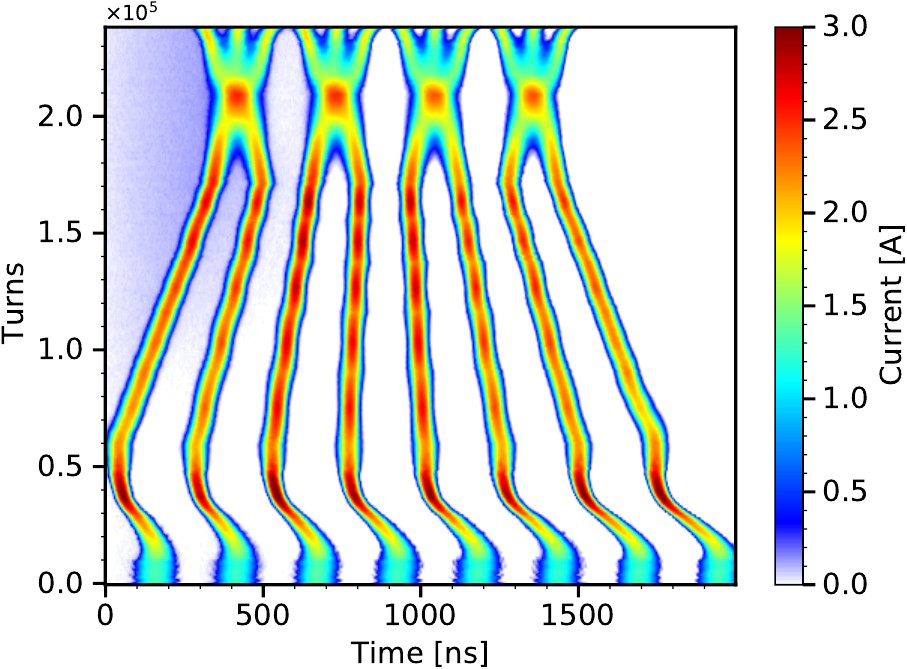}
        \hspace{0.03\textwidth}
        \includegraphics[width=0.45\textwidth]{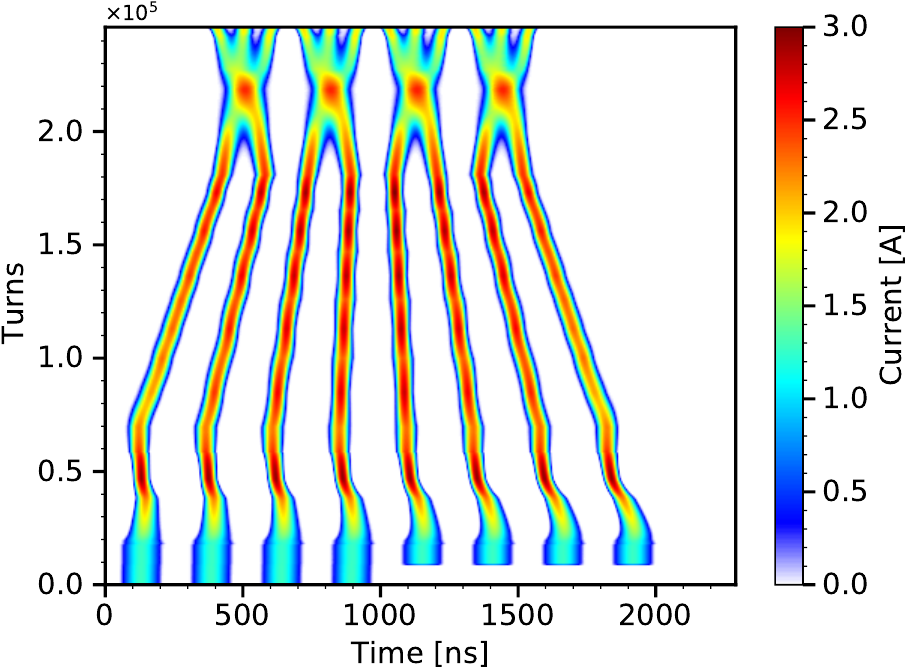}
        \caption{Benchmarking of bunch profiles, shown as waterfall plots, in the PS during RF manipulations. Left: measured profiles, right: simulated profiles. Courtesy of A.~Lasheen.}
        \label{fig:PS-benchmark}
    \end{figure}

    Benchmarking against a known theoretical solution or against measurements is, in accelerator physics, often possible thanks to the availability of well-controlled beam-based measurements; the observables used in simulation and measurement should be made as similar as possible, the same data-analysis methods should be applied to both, and measurement imperfections should, where relevant, be modelled explicitly in the simulation. Figure~\ref{fig:PS-benchmark} shows an example of such a benchmark, comparing measured and simulated bunch profiles during an RF manipulation in the CERN PS.
    
    Finally, continuous-integration (CI) pipelines help maintain trust in a code over time, by ensuring that functionality which worked at one point continues to work as the code evolves. Good CI coverage requires as many unit and integration tests as practical, covering all relevant regions of the code, together with a clear definition of the supported platforms and software versions, and automated checks that the~code still compiles. Tools such as GitHub Actions, GitLab pipelines, Travis, or AppVeyor can be used to trigger this pipeline automatically whenever a new version is deployed, while dedicated coverage tools (e.g.\ codecov or JetBrains dotCover) can report which fraction of the code is actually exercised by the~test suite.

%% file: text/appendix_slippage.tex
\section{Frequency slippage and momentum compaction factor}
\label{app:slippage}

    The frequency slippage experienced by an off-momentum particle is defined with respect to the synchronous particle. The momentum compaction factors $\alpha_i$ and the slippage factors $\eta_i$ introduced in Eq.~\eqref{eq:drift-coord1} are related order by order in $\delta$ through~\cite{Lee2019}
    \begin{equation}
        \frac{\Delta \omega}{\omega_\mathrm{rev}} = \frac{\omega - \omega_\mathrm{rev}}{\omega_\mathrm{rev}} \equiv -\eta(\delta)\,\delta = -(\eta_0 + \eta_1 \delta + \eta_2 \delta^2 + ...)\,\delta \, ,
        \label{eq:slippage-freq}
    \end{equation}
    with
    \begin{align}
        \eta_0 &= \alpha_0 - \frac{1}{\gamma_d^2} = \frac{1}{\gamma_T^2} - \frac{1}{\gamma_d^2} \, , \label{eq:eta0} \\
        \eta_1 &= \frac{3\beta_d^2}{2\gamma_d^2} + \alpha_1 - \alpha_0 \eta_0 \, , \label{eq:eta1} \\
        \eta_2 &= -\frac{\beta_d^2 (5\beta_d^2 - 1)}{2\gamma_d^2} + \alpha_2 - 2\alpha_0\alpha_1 + \frac{\alpha_1}{\gamma_d^2} + \alpha_0^2 \eta_0 - \frac{3\beta_d^2 \alpha_0}{2\gamma_d^2} \, , \label{eq:eta2}
    \end{align}
    where $\gamma_T$ is the transition gamma of the machine. Below transition, i.e.\ for $\gamma_d < \gamma_T$, one has $\eta_0 < 0$, so that particles with higher energy ($\delta > 0$) complete a turn faster ($\Delta\omega > 0$); this ordering reverses above transition.

%% file: text/appendix_symplectic.tex
\section{Derivation of the general symplectic condition}
\label{app:symplectic}

    Consider generic conjugate coordinates $\mathbf{x} \equiv (\mathbf{q}, \mathbf{p}) = (q_1, ..., q_n, p_1, ..., p_n)$, mapped under time evolution into $\mathbf{y} \equiv (\mathbf{Q}, \mathbf{P}) = (Q_1, ..., Q_n, P_1, ..., P_n)$. Defining the matrix
\begin{equation}
    \mathcal{S} \equiv \begin{pmatrix}
        \mathbf{0} & \mathbf{1}\\
        -\mathbf{1} & \mathbf{0}
    \end{pmatrix} \, ,
\end{equation}
Hamilton's equations can be rewritten compactly as
\begin{equation}
    \mathbf{\dot q} = \frac{\partial H}{\partial \mathbf{p}} \, , \qquad
    \mathbf{\dot p} = - \frac{\partial H}{\partial \mathbf{q}} \, ,
    \qquad \Longrightarrow \qquad
    \mathbf{\dot x} = \mathcal{S} \frac{\partial H}{\partial \mathbf{x}} \, .
\end{equation}
Since the new coordinates $\mathbf{y}$ also obey Hamilton's equations,
\begin{equation}
    \mathbf{\dot y} = \mathcal{S} \frac{\partial H}{\partial \mathbf{y}} \, .
\end{equation}
On the other hand, using the mapping $\mathbf{x} \rightarrow \mathbf{y}$, the new coordinates can be expressed in terms of the old ones via the chain rule,
\begin{equation}
    \dot y_i = \frac{d y_i(\mathbf{x})}{d t } = \frac{\partial y_i}{\partial x_k} \frac{d x_k}{d t} = \frac{\partial y_i}{\partial x_k} \mathcal{S}_{kl} \frac{\partial H}{\partial x_l} = \frac{\partial y_i}{\partial x_k} \mathcal{S}_{kl} \frac{\partial y_l}{\partial x_m} \frac{\partial H}{\partial y_m} \, .
\end{equation}
Introducing the mapping, or Jacobian, matrix $\mathcal{M}_{ij} \equiv \partial y_{i}/\partial x_{j}$, this can be written as
\begin{equation}
    \mathbf{\dot y} = \frac{d\mathbf{y}(\mathbf{x})}{d t } = \mathcal{M} \frac{d \mathbf{x}}{d t} = \mathcal{M} \mathcal{S} \frac{\partial H}{\partial \mathbf{x}} = \mathcal{M} \mathcal{S} \mathcal{M}^\mathrm{T} \frac{\partial H}{\partial \mathbf{y}} \stackrel{!}{=} \mathcal{S} \frac{\partial H}{\partial \mathbf{y}} \, .
\end{equation}
Since this must hold for an arbitrary Hamiltonian $H$, the mapping matrix $\mathcal{M}$ has to satisfy the \emph{symplectic condition},
\begin{equation}
    \mathcal{M} \mathcal{S} \mathcal{M}^\mathrm{T} = \mathcal{S} \, ,
    \label{eq:app-symplectic-condition}
\end{equation}
which is equivalent to the mapping preserving the symplectic 2-form,
\begin{equation}
    d \mathbf{p} \wedge d \mathbf{q} \equiv \sum_i d p_i \wedge d q_i = d \mathbf{P} \wedge d \mathbf{Q} \, .
\end{equation}
For a Hamiltonian system with a single pair of conjugate coordinates, as is the case for the longitudinal phase-space coordinates used throughout this paper, Eq.~\eqref{eq:app-symplectic-condition} reduces to the condition $\mathcal{M}^\mathrm{T} \mathcal{S} \mathcal{M} = \mathcal{S}$ used in Section~\ref{sec:longitudinal-tracking}, since $\mathcal{S}$ is then antisymmetric and $2\times2$, so that $\mathcal{M} \mathcal{S} \mathcal{M}^\mathrm{T} = \mathcal{S}$ and $\mathcal{M}^\mathrm{T} \mathcal{S} \mathcal{M} = \mathcal{S}$ are equivalent up to an overall transposition of $\mathcal{S}$ itself, which merely reverses its sign convention.